\documentclass[reprint,aps,prl,superscriptaddress,floatfix,nofootinbib,longbibliography]{revtex4-2}

    \usepackage{graphicx}
	\usepackage{bm}
	\usepackage{amsmath} 
	\usepackage{amssymb}
	\usepackage{braket} 
	\usepackage{epsfig}
	\usepackage{tensor}
	\usepackage{CJKutf8}
	\usepackage[version=4]{mhchem}
	\usepackage{titlesec}
	\usepackage{ifthen}
	\usepackage{sidecap}
	\usepackage{listings}
	\usepackage{array}
	\usepackage{floatrow}
	\usepackage{multirow}
	\usepackage{enumitem}
	\usepackage[para,online,flushleft]{threeparttablex}
	\usepackage{float}
	\floatstyle{plaintop}
	\restylefloat{table}
	\usepackage{booktabs}
    \usepackage{mathtools}
    \usepackage{xcolor}
    \usepackage{changes}

	\usepackage{hyperref}
	\hypersetup{breaklinks=true,colorlinks=true,linkcolor=blue,citecolor=blue,filecolor=magenta,urlcolor=cyan}
	
	\usepackage[all]{hypcap}

	\usepackage{xcolor}
	\definecolor{pastelgray}{rgb}{0.81, 0.81, 0.77}
	\definecolor{beaublue}{rgb}{0.9, 0.9, 0.93}

	\usepackage{orcidlink}
	\usepackage{tikz,xcolor,hyperref}
	
	\definecolor{lime}{HTML}{A6CE39}
	\DeclareRobustCommand{\orcidicon}{
		\begin{tikzpicture}
			\draw[lime, fill=lime] (0,0) 
			circle [radius=0.16] 
			node[white] {{\fontfamily{qag}\selectfont \tiny ID}};
			\draw[white, fill=white] (-0.0625,0.095) 
			circle [radius=0.007];
		\end{tikzpicture}
		\hspace{-2mm}
	}
	
	\foreach \x in {A, ..., Z}{%
		\expandafter\xdef\csname orcid\x\endcsname{\noexpand\href{https://orcid.org/\csname orcidauthor\x\endcsname}{\noexpand\orcidicon}}
	}

	\makeatletter
	\def\@bibdataout@aps{%
		\immediate\write\@bibdataout{%
			@CONTROL{%
				apsrev41Control%
				\longbibliography@sw{%
					,author="08",editor="1",pages="1",title="0",year="1"%
				}{%
					,author="08",editor="1",pages="1",title="",year="1"%
				}%
			}%
		}%
		\if@filesw \immediate \write \@auxout {\string \citation {apsrev41Control}}\fi
	}
	\makeatother

	\newcolumntype{Y}{>{\centering\arraybackslash}X}

\begin{document}
		\begin{CJK*}{UTF8}{gbsn}
			
			\title{
            Spin Teleportation via Bell-Triplet States Emergent from Proton-Proton Scattering
            }

		\author{Z.~X.~Shen}
		\affiliation{Key Laboratory of Nuclear Physics and Ion-beam Application (MOE), Institute of Modern Physics, Fudan University, Shanghai 200433, China}
		\affiliation{Shanghai Research Center for Theoretical Nuclear Physics, NSFC and Fudan University, Shanghai 200438, China}
		
		\author{H.~Y.~Shang}
	\affiliation{Key Laboratory of Nuclear Physics and Ion-beam Application (MOE), Institute of Modern Physics, Fudan University, Shanghai 200433, China}
	\affiliation{Shanghai Research Center for Theoretical Nuclear Physics, NSFC and Fudan University, Shanghai 200438, China}
	
	\author{Y.~G.~Ma\,\orcidlink{0000-0002-0233-9900}}\email{Email: mayugang@fudan.edu.cn}
	\affiliation{Key Laboratory of Nuclear Physics and Ion-beam Application (MOE), Institute of Modern Physics, Fudan University, Shanghai 200433, China}
	\affiliation{Shanghai Research Center for Theoretical Nuclear Physics, NSFC and Fudan University, Shanghai 200438, China}
    \affiliation{School of Physics, East China Normal University, Shanghai 200241, China}
	
	\author{D.~Bai\,\orcidlink{0000-0001-7116-721X}}\email{Email: dbai@hhu.edu.cn}
	\affiliation{College of Mechanics and Engineering Science, Hohai University, Nanjing 211100, China}
	
	\author{S.~M.~Wang\,\orcidlink{0000-0002-8902-6842}}\email{Email: wangsimin@fudan.edu.cn}
	\affiliation{Key Laboratory of Nuclear Physics and Ion-beam Application (MOE), Institute of Modern Physics, Fudan University, Shanghai 200433, China}
	\affiliation{Shanghai Research Center for Theoretical Nuclear Physics, NSFC and Fudan University, Shanghai 200438, China}
		
	\author{Z.~C.~Xu}
	\affiliation{Key Laboratory of Nuclear Physics and Ion-beam Application (MOE), Institute of Modern Physics, Fudan University, Shanghai 200433, China}
	\affiliation{Shanghai Research Center for Theoretical Nuclear Physics, NSFC and Fudan University, Shanghai 200438, China}

    \author{Y.~Ayyad\,\orcidlink{0000-0001-8604-4976}}\email{Email: yassid.ayyad@usc.es}
	\affiliation{IGFAE, Universidade de Santiago de Compostela, E-15782, Santiago de Compostela, Spain}

       \author{C.~Filgueira}
	\affiliation{IGFAE, Universidade de Santiago de Compostela, E-15782, Santiago de Compostela, Spain}

	\begin{abstract}

Entanglement is a key resource in quantum information science, yet its properties and applications in nuclear systems remain largely unexplored. Here, using proton-proton scattering as a quantum laboratory, we report the emergence of a near-pure Bell-triplet state at a laboratory energy of 151 MeV and a center-of-mass scattering angle of 90 degrees. In this unique kinematic regime, the scattering amplitude functions as a transition operator connecting distinct Bell states. Building upon this emergent resource, we propose a quantum teleportation protocol for proton spins, exploiting the intrinsic Hamiltonian of the strong interaction to perform the requisite Bell measurement effectively. These findings effectively bridge few-body nuclear physics and quantum technology, establishing proton-proton scattering as both a source of high-fidelity entanglement and a natural processor for quantum information.

	\end{abstract}
	\maketitle
	\end{CJK*}

\textit{Introduction}---Quantum entanglement is a curious yet fundamental resource of nature, characterized by strong nonlocal correlations with no classical description \cite{Horodecki2009}. Harnessing this property is crucial for next-generation technologies ranging from quantum computing to quantum sensing \cite{Nielsen_Chuang_2010}. While entanglement is routinely engineered in controlled systems like trapped ions \cite{Cirac1995,Monroe1995,Molmer1999}, ultracold atoms \cite{Sorensen2001,Athreya2026,Yang2020}, and superconducting circuits \cite{Devoret1984,Martinis1985,Devoret1985}, it is also an inherent and ubiquitous feature across vastly different scales, from the subatomic world to theories of quantum gravity \cite{BESIII2019,BESIII2022,ATLAS2024,Chen2024,BESIII2025,Aziz2025}.  

In this broader context, 
subatomic systems with rich spin-dependent dynamics offer a natural arena for producing and manipulating spin-entangled states, with recent efforts probing entanglement and Bell nonlocality in high-energy hadronic and electroweak reactions \cite{Hentschinski2023,ATLAS2024,Guo2025,Ehataht2024,Afik2025}.
Alongside these efforts, which also highlight the challenge of realizing loophole-free Bell tests at colliders \cite{Abel1992,Bechtle2025,Abel2025,Li2024,Fabbrichesi2025,Low2025,Demina2025,Barr2024}, the nuclear domain serves as a highly complementary platform for the generation and control of spin entanglement.
This quantum-information perspective not only yields novel insights into nuclear forces, structure, and dynamics \cite{Beane2019,Johnson2019,Robin2020,Bai2022,Kirchner2023,Ma2023,Robin2024,Qiang2024,Shang2025,Robin2025,Hentschinski2023,Afik2025,Guo2025,Ehataht2024,Pacheco2025,Shin2019,Rosenfeld2017,Gu2025,Lin2025}, but also drives innovations in experimental techniques for probing quantum correlations in nuclei \cite{Bai2023a}.

As one of the most fundamental processes in nuclear physics, nuclear scattering serves as a primary tool for probing nuclear forces and structure \cite{Epelbaum2008,Machleidt2011}. 
Interest in spin entanglement in this setting dates back to the pioneering nuclear Bell test \cite{LamehiRachti1976}, which exploited the spin-singlet state generated by low-energy proton-proton ($pp$) scattering. Owing to technical challenges, the field then entered a long hiatus, and a second Bell test using proton singlet pairs was not realized until 2006 \cite{Sakai2006}. More recently, this direction has seen substantial theoretical progress, with several studies investigating the spin entanglement properties of $np$, $nn$, and neutron-deuteron scattering using the full $\mathbb{S}$-matrix formalism \cite{Bai2023,Miller2023,Cavallin2025,Witala2025,Witala2025_2}. In contrast, a parallel comprehensive investigation for the $pp$ scattering---the experimentally most accessible two-nucleon channel---has remained incomplete, leaving a conspicuous gap in our understanding of spin entanglement across all nucleon-nucleon sectors.

In this work, we complement these efforts by focusing on $pp$ scattering and show that it reveals physics not exposed by conventional observables alone. Although modern interactions reproduce $pp$ differential cross sections up to $\sim 350$~MeV with $\chi^2/\mathrm{dof}\!\sim\!1$ \cite{Stoks1993,Wiringa1995,Machleidt2001,Epelbaum2014,Entem2017}, this apparent completeness does not extend to entanglement-sensitive structure. By adopting entanglement measures as quantum-information-inspired probes, we identify a unique, previously unreported kinematic region characterized by the emergence of a nearly pure Bell-triplet state. This window provides a new benchmark for constraining nucleon–nucleon interactions and a realistic venue for Bell-type tests. Moreover, by tracing the dynamical origin of this phenomenon, we find that the spin amplitude effectively acts as a Bell transition operator, positioning nuclear scattering as a promising resource for quantum information processing.

As a concrete application enabled by this discovery, we propose a quantum teleportation protocol \cite{Bennett1993b} for proton spins, driven directly by the intrinsic Hamiltonian of the strong interaction. In the identified kinematic window, the spin amplitude effectively performs the requisite Bell measurement, enabling intact transfer of an unknown spin state between protons without externally engineered two-qubit gates.
In this sense, teleportation is not merely a conceptual analogy but a novel experimental measurement modality---one that leverages entanglement generated by realistic nuclear forces to realize quantum-information tasks in the femtometer/MeV domain.

\textit{Spin amplitude and entanglement measures}---The two-proton system is described within the combined momentum–spin Hilbert space, $\mathcal{H}=\mathcal{H}_p\otimes\mathcal{H}_s$, where $\mathcal{H}_s=\mathcal{H}_A\otimes\mathcal{H}_B$ denotes the four-dimensional spin space for two spin-$\tfrac{1}{2}$ particles. With an initial spin density matrix $\rho_{i}$, the final state is $\rho_{f}={M\rho_{i}M^{\dagger}}/{\operatorname{Tr}(M\rho_{i}M^{\dagger})}$. The spin amplitude $M$, which is a $4\times4$ matrix encoding the complete spin dependence of $pp$ scattering, can be evaluated via the Nijmegen PWA93 database \cite{NNOnline,Stoks1993} and chiral effective field theory ($\chi$EFT) \cite{Saha2023}. The Coulomb correction is incorporated following the treatment in Refs.\ \cite{Stoks1993,Bergervoet1988}. In all calculations, the $z$-axis is aligned with the initial relative momentum.

	\begin{figure}[t]
		\centering
		\includegraphics[width=\linewidth]{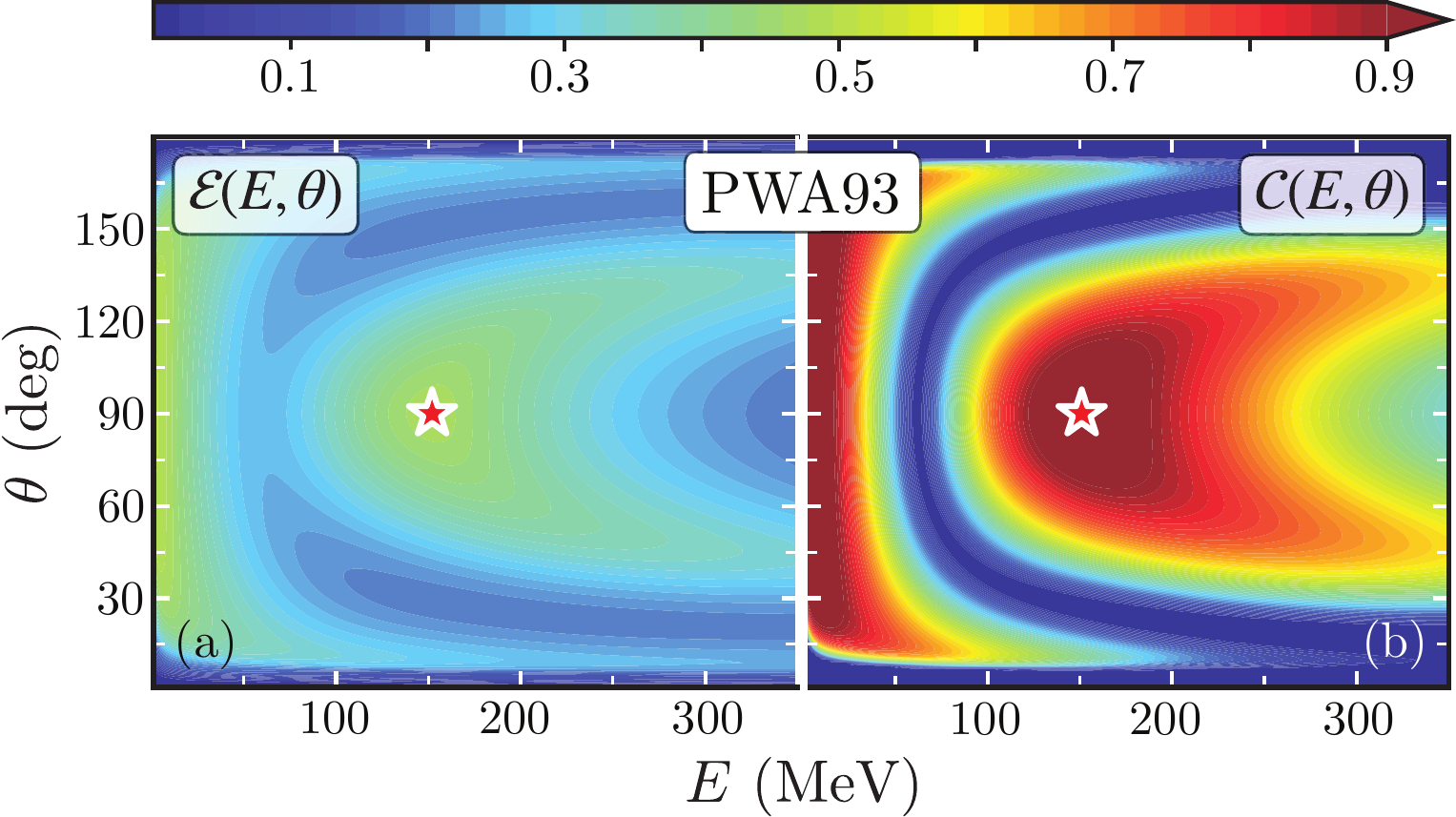}
		\caption{(a) Entanglement power $\mathcal{E}$ and (b) concurrence $\mathcal{C}$ for $pp$ scattering, shown as functions of the laboratory kinetic energy $E$ and the center–of–mass angle $\theta$. Both quantities are calculated from the Nijmegen PWA93 database, with the concurrence evaluated for a completely unpolarized initial state. The red pentagrams mark the second local maxima in each distribution at $(E,\theta)=(151\ \text{MeV}, 90^\circ)$.}
		\label{fig:one}
	\end{figure}
   
To quantify the spin entanglement properties of $pp$ scattering, we employ entanglement power and concurrence. For a scattering event at laboratory kinetic energy $E$ and center-of-mass angle $\theta$, the entanglement power $\mathcal{E}(E,\theta)$ quantifies the average entanglement generated by the scattering operator acting on all possible separable initial spin states. These initial states are parameterized as
\begin{equation*}
|\chi_{i}\rangle = \begin{bmatrix} \cos\frac{\theta_1}{2}, \ e^{i\varphi_1}\sin\frac{\theta_1}{2} \end{bmatrix}^{\mathsf{T}} \otimes \begin{bmatrix} \cos\frac{\theta_2}{2}, \ e^{i\varphi_2}\sin\frac{\theta_2}{2} \end{bmatrix}^{\mathsf{T}},
\end{equation*}
where the angles define the initial spin orientations on the Bloch spheres of the two protons \cite{Zanardi2000_v2,Datta2005}. The corresponding final state is $\ket{\chi_{f}}=M\ket{\chi_{i}}/\sqrt{\bra{\chi_{i}}M^\dagger M\ket{\chi_{i}}}$. Consequently, the entanglement power is given by
\begin{equation}
\mathcal{E}(E,\theta)=1-\int\!\frac{d\Omega_1}{4\pi}\frac{d\Omega_2}{4\pi}\operatorname{Tr}[\rho_1^2],
\end{equation}
where $\rho_1=\operatorname{Tr}_2(\ket{\chi_{f}}\bra{\chi_{f}})$ is the reduced density matrix for one proton, and the integral runs over all spin orientations. 

The concurrence $\mathcal{C}$ quantifies the entanglement of a two-qubit mixed state $\rho$. It is defined as 
\begin{equation}
\mathcal{C}(\rho)=\max\{0,\lambda_1-\lambda_2-\lambda_3-\lambda_4\},
\end{equation}
where $\{\lambda_i\}$ are the square roots of the eigenvalues of $R=\rho(\sigma_y\!\otimes\!\sigma_y)\rho^*(\sigma_y\!\otimes\!\sigma_y)$ in descending order \cite{Hill1997,Wootters1998}.
This measure ranges from 0 to 1, where 0 indicates an unentangled (separable) state and 1 corresponds to a maximally entangled state. For instance, all four Bell states yield $\mathcal{C}=1$, confirming their maximal entanglement. As an experimentally accessible measure, concurrence is widely used as a genuine entanglement quantifier for two qubits \cite{Wootters1998}. 
	
\textit{Emergence of the entangled spin-triplet state}---The entanglement power extracted from the Nijmegen PWA93 database (including Coulomb effects) is shown in Fig.\,\ref{fig:one}(a). It exhibits a symmetric profile at the scattering angle $\theta=90^\circ$, reflecting the indistinguishability of the two protons. Two pronounced enhancements are observed: one at low laboratory energies ($E<10$~MeV) and another around $(E,\theta)=(151~\text{MeV},90^\circ)$, hereafter denoted as $(E_\odot,\theta_\odot)$. 
These regions correspond to kinematic conditions that favor strong spin entanglement. The well-known low-energy enhancement originates from the ${}^1S_0$ partial-wave scattering, which generates the antisymmetric singlet state $|\Psi^{-}\rangle = (\ket{\uparrow\downarrow}-\ket{\downarrow\uparrow})/\sqrt{2}$ \cite{Glockle1983,Stoks1993}, long recognized as the essential configuration for nuclear Bell tests with proton singlet pairs \cite{LamehiRachti1976,Sakai2006}.
In sharp contrast, the distinct peak at $(E_\odot,\theta_\odot)$ reveals a previously unexplored regime where the scattering process gives rise to a strongly entangled spin-triplet state.

\begin{figure}[t]
\centering
\includegraphics[width=\linewidth]{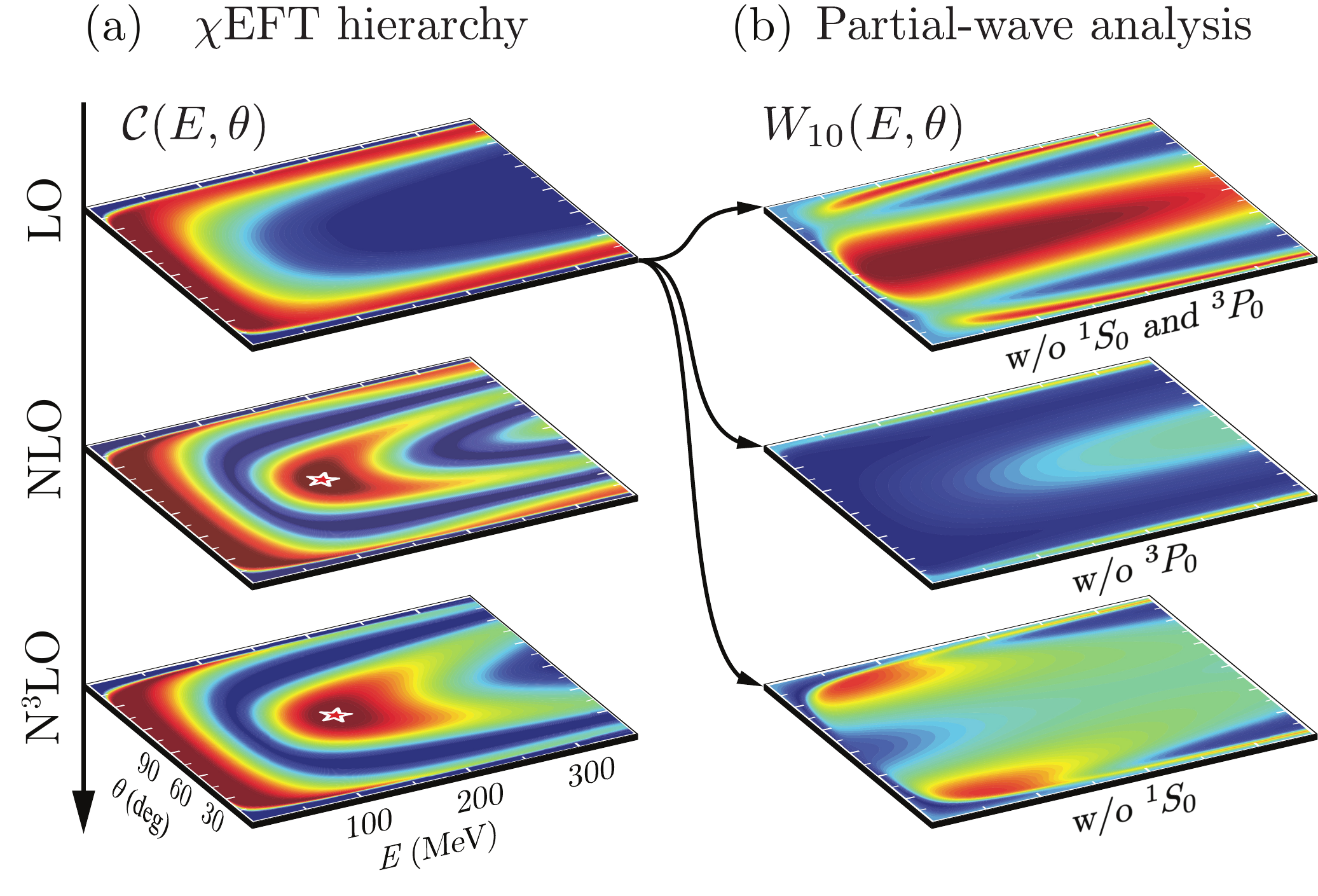}
\caption{(a) Concurrence $\mathcal{C}$ calculated using chiral EFT at different orders. (b) Partial-wave decomposition of the Bell-triplet weight $W_{10}$ at LO in chiral EFT. The three panels, from top to bottom, respectively show the results excluding the contributions from $^1S_0$ and $^3P_0$ components.}
\label{fig:two}
\end{figure}
    
While the concept of entanglement power has garnered theoretical interest in nuclear physics \cite{Bai2023,Miller2023,Cavallin2025,Witala2025}, its direct experimental determination remains challenging. Given this limitation, we adopt concurrence $\mathcal{C}$ as a more practical and experimentally accessible measure of spin entanglement. The concurrence map for a completely unpolarized initial state, Fig.\,\ref{fig:one}(b), closely mirrors the entanglement-power landscape and reaches $\mathcal{C}=0.977$ at $(E_\odot,\theta_\odot)$. At this kinematic point, the reconstructed final-state density matrix is dominated by a single Bell-triplet component,
\begin{equation}
\rho_f = a_1 |\Psi^+\rangle\langle\Psi^+| + a_2 |\Phi^-\rangle\langle\Phi^-| + a_3 |\Psi^-\rangle\langle\Psi^-|,
\end{equation}
with $a_1 = 0.988$, $a_2 = 0.009$, and $a_3 = 0.003$. Here $\ket{\Psi^\pm}=(\ket{\uparrow\downarrow}\pm\ket{\downarrow\uparrow})/\sqrt{2}$ and $\ket{\Phi^\pm}=(\ket{\uparrow\uparrow}\pm\ket{\downarrow\downarrow})/\sqrt{2}$ are the Bell basis states. 
To an excellent approximation, $pp$ scattering in the vicinity of $(E_\odot,\theta_\odot)$ produces the pure Bell-triplet state $\ket{\Psi^+}$, providing a high-quality triplet source for further Bell tests and related applications (see Supplemental Material~\cite{SM} for additional discussion).
\nocite{Horodecki1995,Brunner2014,Clauser1969,
Comparat1974,McNaughton1985,Meyer1983,
Riebe2004,Barrett2004,Steffen2013,Nolleke2013,
Kox1993,Mohr2025,Bargmann1959,Jackson1999,
YuEberly2006,PDG2024,LynchDahl1991,Leo1994,
Carlson1996}

\textit{Nature of the Bell-triplet state}---%
To uncover the origin of the Bell-triplet window, we compute the concurrence using chiral EFT interactions from leading order (LO) up to N$^3$LO \cite{Saha2023}, with Coulomb effects incorporated as in Refs.\ \cite{Stoks1993,Bergervoet1988}. Along $\theta =90^\circ$, Fig.~\ref{fig:two}(a) shows that realistic interactions beyond LO reproduce the pronounced concurrence peak near $E_{\rm lab}\simeq151$~MeV, while LO largely misses it.
As demonstrated below, the Bell-triplet window is actually a delicate consequence of the interplay between two ingredients: (i) the phase-shift pattern, through the resulting partial-wave balance, which largely fixes the energy where such a window can occur; and (ii) a microscopic spin-mixing mechanism that drives the outgoing state toward an almost pure triplet.

Indeed, the concurrence enhancement requires the dominant ${}^1S_0$ and ${}^3P_0$ contributions to be reduced so that other amplitudes can interfere coherently. We illustrate this by evaluating the Bell-triplet weight
$W_{10}=\mathrm{Tr}\!\left(\rho_f\,\ket{\Psi^+}\!\bra{\Psi^+}\right)$
at LO while selectively excluding the ${}^1S_0$ and/or ${}^3P_0$ components. As shown in Fig.~\ref{fig:two}(b), removing these channels opens a broad triplet-favored region near $\theta =90^\circ$, consistent with the interpretation that the phase shifts primarily determine where the window can appear.

Furthermore, within the N$^3$LO chiral EFT, we scale the tensor component of the nuclear potential by $c_T$ (i.e., $V_T\!\to c_TV_T$) and recompute the entanglement measures. Reducing $c_T$ strongly suppresses the concurrence peak, while restoring the tensor strength recovers both $\mathcal{C}$ and $W_{10}$ near $(E_\odot,\theta_\odot)$. This behavior shows that the tensor operator
$S_{12}=3(\boldsymbol{\sigma}_{1}\cdot\hat{\boldsymbol r})(\boldsymbol{\sigma}_{2}\cdot\hat{\boldsymbol r})-\boldsymbol{\sigma}_{1}\!\cdot\!\boldsymbol{\sigma}_{2}$
provides a mechanism for coupling different magnetic components and transferring amplitude into the $m_s=0$ triplet sector. The tensor contribution alone, however, does not exhaust the mechanism. As shown in the Supplemental Material based on the Argonne series of nuclear forces~\cite{SM}, a central interaction does not produce the triplet window, a tensor-only extension gives only a partial enhancement, and the inclusion of spin-orbit operators brings the result close to the realistic case~\cite{Wiringa1995,Wiringa2002}. Precision measurements of spin entanglement would consequently provide a direct and complementary constraint on the spin-dependent nuclear interaction, especially its tensor and spin-orbit sectors.

\begin{figure}[t]
\centering
\includegraphics[width=\linewidth]{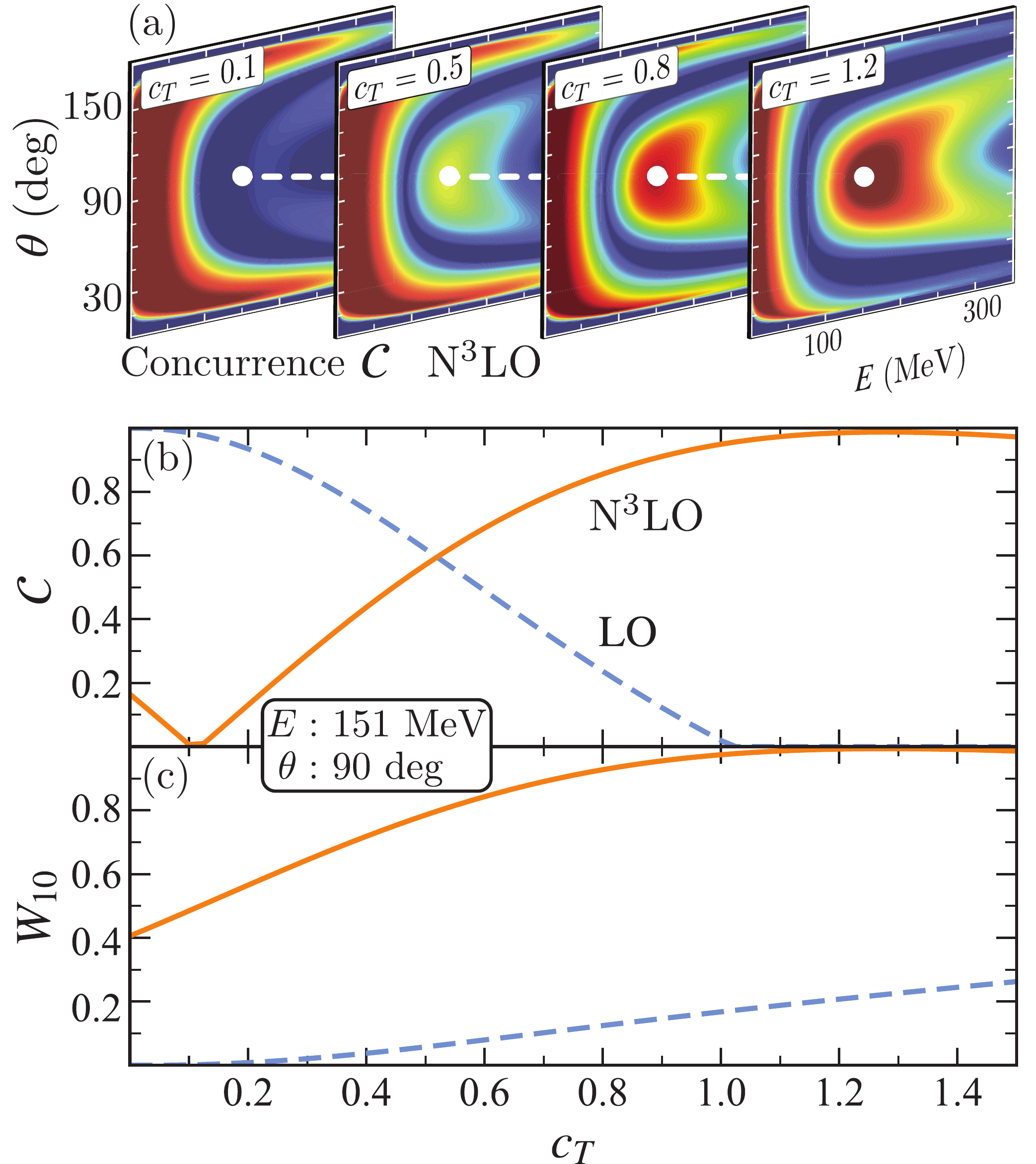}
\caption{(a) Concurrence $\mathcal{C}$ calculated using the N$^3$LO chiral EFT interaction with the tensor strength $c_T$ varied. (b) Concurrence $\mathcal{C}$ and (c) Bell-triplet weight $W_{10}$ at $(E,\theta)=(151\ \text{MeV}, 90^\circ)$, calculated using both LO and N$^3$LO chiral EFT interactions, plotted as a function of $c_T$.}
\label{fig:three}
\end{figure}

\textit{Bell transition operator}---%
The Bell-triplet dominance around $(E_\odot,\theta_\odot)$ is accompanied by a remarkably simple structure at the amplitude level. In the Bell basis, the spin amplitude is dominated by a single matrix element,
\begin{equation}
M(E_\odot,\theta_\odot)\simeq (-3.845-i0.058)\ket{\Psi^+}\!\bra{\Phi^-},
\end{equation}
with all other Bell-basis components smaller by more than one order of magnitude. Consequently, for any incoming two-proton state $\rho_i$ with nonzero overlap with $\ket{\Phi^-}$,
\begin{equation}
\rho_f \propto M\rho_i M^\dagger \;\Rightarrow\; \rho_f \simeq \ket{\Psi^+}\!\bra{\Psi^+},
\end{equation}
showing that $pp$ scattering near $(E_\odot,\theta_\odot)$ coherently converts the $\ket{\Phi^-}$ component of the input into the Bell triplet.

\begin{figure*}[!t]
	\centering
	\includegraphics[width=0.95\linewidth]{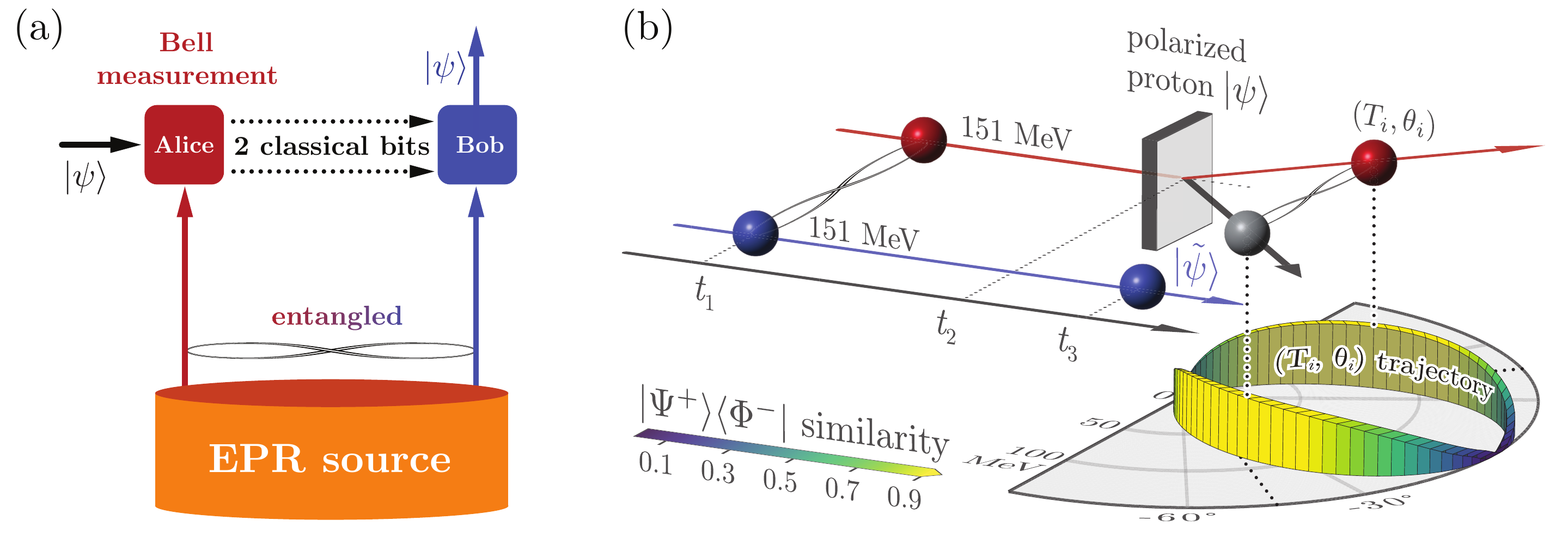}
		\caption{(a) Standard quantum teleportation protocol using shared entanglement and Bell measurement. (b) Proposed proton spin teleportation at 151 MeV. Scattering near $90^\circ$ ($t_2$) realizes approximately a Bell transition operator, teleporting the target spin state $|\psi\rangle$---up to an inversion of the quantization axis---to the remote proton ($t_3$). The pie plot shows the kinematically allowed $(T_i,\theta_i)$ locus (laboratory kinetic energy and scattering angle) for a scattered proton; the histogram height and color encode the similarity $\mathcal{S}$ \cite{WatrousTQI} between the spin amplitude $M$ and the Bell transition operator $|\Psi^+\rangle\langle\Phi^-|$ (PWA93), indicating a broad angular range where the transition is effective (see Supplemental Material \cite{SM} for additional discussion).}
	\label{fig:qt}
\end{figure*}

In quantum-information terms, $M$ realizes approximately a Bell transition operator, mapping $\ket{\Phi^-}$ into a different Bell state $\ket{\Psi^+}$. This behavior contrasts with the low-energy regime ($E\!\lesssim\!10$~MeV), where $M\propto\ket{\Psi^-}\!\bra{\Psi^-}$ acts as an effective Bell-singlet projector. The evolution from projective to transition-operator-like behavior as the energy increases reveals that the nuclear spin amplitude can function as a physically realized spin-entanglement processor, with the strong interaction providing the underlying mechanism for Bell-state manipulation. By effectively implementing a Bell measurement through this transition mechanism, the $pp$ scattering provides the physical resource required for MeV-scale spin teleportation discussed below.

\textit{Spin teleportation}---Quantum teleportation is well established in photonic and atomic platforms \cite{Pirandola2015,Hu2023}, but it has not been demonstrated in MeV-scale nuclear reactions, despite early proposals based on standard nuclear scattering and polarization techniques \cite{Kostenko2002,Ivanov2023}. Bridging this gap would extend quantum information science into the femtometer domain and ultrafast dynamics governed by the strong interaction, with reduced exposure to external environmental noise. Moreover, it could provide a pathway to map nuclear states onto accessible platforms, unlocking new avenues for probing nuclear structure via quantum state tomography. 

While the standard protocol requires an entangled source, a Bell measurement, and classical communication as shown in Fig.~\ref{fig:qt}(a), the nuclear environment presents unique opportunities and challenges.  
Entangled proton pairs can be generated through elastic $pp$ scattering~\cite{LamehiRachti1976}, charge-exchange reactions like $p(d,{}^2\text{He})n$~\cite{Sakai2006}, and $2p$ decay of ${}^6\text{Be}$~\cite{Wang2021,Oishi2025}. This makes the entangled resource experimentally accessible with standard beam/target and polarimetry techniques.
However, unlike standard gate-based teleportation, which relies on programmable inverse Bell-basis transformations via Hadamard and controlled-NOT gates, such explicit gate-level control is very difficult for individual hadronic spin qubits in a MeV proton-scattering experiment.
To overcome this, we propose harnessing the intrinsic scattering dynamics. The $pp$ spin amplitude offers two avenues: the low-energy Bell-singlet projector $M\propto\ket{\Psi^-}\bra{\Psi^-}$ ($E<10$ MeV), and the newly identified Bell transition operator, $M\propto\ket{\Psi^+}\bra{\Phi^-}$, emerging around $(E_\odot,\theta_\odot)$. Below, we focus on the second avenue; the low-energy singlet-based scheme is detailed in the Supplemental Material~\cite{SM}.

A schematic is shown in Fig.~\ref{fig:qt}(b). Because all participants are protons (identical fermions), we use the indices and colors in Fig.~\ref{fig:qt}(b) to label \emph{spatially separated modes} (arms/trajectories) rather than intrinsic particle identity; in the postselected subspace with one proton in each mode, the spin space is isomorphic to that of three distinguishable qubits. 
At time $t_1$, an entangled Bell-singlet pair (red/blue) can be produced by the $p(d,{}^2\text{He})n$ reaction~\cite{Sakai2006} or other established mechanisms discussed above---and split into two arms: a high-energy proton (red, mode~2) tuned to $E\simeq151$~MeV and a remote partner (blue, mode~3). 
The proton in mode~2 then scatters elastically from a polarized proton target (gray, mode~1) prepared in the state $\ket{\psi}$. In the Bell-triplet window ($E\simeq151$~MeV and $\theta\simeq90^\circ$), the spin amplitude is well approximated by a Bell-state transition, $M\propto\ket{\Psi^+}\!\bra{\Phi^-}$, which transfers the target spin information to the remote proton. We postselect coincidence events in which the two outgoing protons populate two detector arms around $\pm45^\circ$ in the laboratory frame, thereby defining the two-qubit output modes. Consequently, at $t_3$, the blue proton carries state $\ket{\tilde{\psi}}$, which reproduces $\ket{\psi}$ up to a quantization axis inversion (i.e., a fixed, correctable single-qubit operation). The polarization of the blue proton can then be read out using standard secondary-scattering proton polarimetry~\cite{LamehiRachti1976,Sakai2006} and compared with the known target polarization, providing an experimental verification of the transferred spin state. 
Experimental spin transport, decoherence, and receiver-proton spin readout, including estimates of the detection efficiency, are discussed in the Supplemental Material~\cite{SM}.
The extension of the protocol from a pure state to a general mixed state is also given there.

Central to this scheme is the dynamical emergence of the Bell transition operator $\ket{\Psi^+}\bra{\Phi^-}$ near $(E_\odot,\theta_\odot)$. We quantify this by calculating the similarity between the physical spin amplitude $M$ (from PWA93) and the Bell transition operator $\ket{\Psi^+}\bra{\Phi^-}$ using the normalized trace metric $\mathcal{S}(E,\theta)
=
{\big|\mathrm{Tr}\!\left(M^\dagger(E,\theta)\,\ket{\Psi^+}\bra{\Phi^-}\right)\big|}
/{\sqrt{\mathrm{Tr}\!\left(M^\dagger(E,\theta)\,M(E,\theta)\right)}}$ \cite{WatrousTQI}.
As shown in Fig.~\ref{fig:qt}(b) and Supplemental Material~\cite{SM}, a high-similarity ``island" appears around $(E_\odot,\theta_\odot)$, indicating that the nuclear force naturally generates an amplitude close to the target Bell-state transition over a broad kinematic window.

Regarding the classical communication required for spin teleportation, kinematic postselection makes the scattering between the red and gray protons in Fig.~\ref{fig:qt}(b) an intrinsically probabilistic switch for the occurrence of spin teleportation. As a result, the usual two-bit message (required to identify one of four Bell outcomes) is effectively reduced to a single bit: whether a valid event occurred at the prescribed kinematics. This ``event-ready" signal is provided by coincident detection of the outgoing protons, which both heralds successful operation and explicitly reflects the probabilistic nature of our protocol.

\textit{Summary}---Our study exemplifies how concepts from quantum information science can provide new perspectives on nucleon-nucleon scattering and nuclear force, long regarded as well understood. Specifically, we reveal a pronounced region around $(E_\odot,\theta_\odot)=(151\ \text{MeV},90^\circ)$ in $pp$ scattering that is highly conducive to spin entanglement generation. Calculations using the Nijmegen PWA93 database and chiral EFT interactions demonstrate that the scattering output at this kinematic point forms a nearly pure Bell-triplet state $\ket{\Psi^+}$ for all chiral orders beyond LO. Analysis of the underlying spin amplitude further reveals that it acts predominantly as a Bell transition operator of the form $\propto\ket{\Psi^+}\bra{\Phi^-}$. Building on this finding, we propose a proton-spin teleportation protocol in which the $pp$ spin amplitude mimics the Bell measurement. The realization of such spin teleportation via the strong interaction would extend the reach of quantum information protocols into the femtometer regime, offering a novel method for the faithful transfer and manipulation of nuclear quantum states.

\textit{Acknowledgments}---We thank Dean Lee, Mart Rentmeester and Kentaro Yako for helpful communications and discussions. This work is supported by the National Natural Science Foundation of China (Grants No.\ 12347106, No.\ 12547102, No.\ 12447122, and No.\ 12375122), the National Key Research and Development Program of China (MOST 2023YFA1606404 and MOST 2022YFA1602303), and the China Postdoctoral Science Foundation (Grant No.\ 2024M760489).
During the preparation of this manuscript, the authors used large language models to assist with specific aspects of code development, mathematical derivations, and language editing. The AI-generated outputs were independently verified and audited by the authors, who take full responsibility for the scientific content of the manuscript.

\textit{Data availability}---The numerical data supporting the findings of this article are available from the corresponding authors upon reasonable request.

\textit{Note added}---This work on spin teleportation was initiated in October 2025. The main findings were presented by D.B.\ on behalf of the collaboration at the Solenoidal Spectrometer Workshop 2025 on December 9; slides are available at \cite{Bai2025_Workshop}. 

%

	\end{document}


\begin{CJK*}{UTF8}{gbsn}
		
		\title{
			Supplemental Material for ``Spin Teleportation via Bell-Triplet States Emergent from Proton-Proton Scattering''
		}
		
		\author{Z.~X.~Shen}
		\affiliation{Key Laboratory of Nuclear Physics and Ion-beam Application (MOE), Institute of Modern Physics, Fudan University, Shanghai 200433, China}
		\affiliation{Shanghai Research Center for Theoretical Nuclear Physics, NSFC and Fudan University, Shanghai 200438, China}
		
		\author{H.~Y.~Shang}
		\affiliation{Key Laboratory of Nuclear Physics and Ion-beam Application (MOE), Institute of Modern Physics, Fudan University, Shanghai 200433, China}
		\affiliation{Shanghai Research Center for Theoretical Nuclear Physics, NSFC and Fudan University, Shanghai 200438, China}
		
		\author{Y.~G.~Ma\,\orcidlink{0000-0002-0233-9900}}\email{Email: mayugang@fudan.edu.cn}
	\affiliation{Key Laboratory of Nuclear Physics and Ion-beam Application (MOE), Institute of Modern Physics, Fudan University, Shanghai 200433, China}
	\affiliation{Shanghai Research Center for Theoretical Nuclear Physics, NSFC and Fudan University, Shanghai 200438, China}
	\affiliation{School of Physics, East China Normal University, Shanghai 200241, China}
	
	\author{D.~Bai\,\orcidlink{0000-0001-7116-721X}}\email{Email: dbai@hhu.edu.cn}
	\affiliation{College of Mechanics and Engineering Science, Hohai University, Nanjing 211100, China}

	\author{S.~M.~Wang\,\orcidlink{0000-0002-8902-6842}}\email{Email: wangsimin@fudan.edu.cn}
	\affiliation{Key Laboratory of Nuclear Physics and Ion-beam Application (MOE), Institute of Modern Physics, Fudan University, Shanghai 200433, China}
	\affiliation{Shanghai Research Center for Theoretical Nuclear Physics, NSFC and Fudan University, Shanghai 200438, China}

	\author{Z.~C.~Xu}
	\affiliation{Key Laboratory of Nuclear Physics and Ion-beam Application (MOE), Institute of Modern Physics, Fudan University, Shanghai 200433, China}
	\affiliation{Shanghai Research Center for Theoretical Nuclear Physics, NSFC and Fudan University, Shanghai 200438, China}
	
	\author{Y.~Ayyad\,\orcidlink{0000-0001-8604-4976}}\email{Email: yassid.ayyad@usc.es}
	\affiliation{IGFAE, Universidade de Santiago de Compostela, E-15782, Santiago de Compostela, Spain}
	
	\author{C.~Filgueira}
	\affiliation{IGFAE, Universidade de Santiago de Compostela, E-15782, Santiago de Compostela, Spain}
	
	\begin{abstract}
		\bigskip
		\noindent
		This Supplemental Material contains:
		\begin{itemize}[leftmargin=*,itemsep=2pt,topsep=2pt]
			\item \textbf{Triplet source and feasibility estimates:} an order-of-magnitude production-rate estimate for near-pure Bell-triplet proton pairs from unpolarized $pp$ scattering at $(E_\odot,\theta_\odot)$, based on the differential cross section and typical beam/target/acceptance parameters.
			\item \textbf{Nuclear Bell tests in the triplet regime:} an assessment of CHSH-inequality violations using the Horodecki criterion, together with the corresponding spin-correlation function $F(\bm n_1,\bm n_2)$ evaluated from the PWA93 spin-density matrix. We also discuss the detection and locality loopholes.
			\item \textbf{Realistic interactions and operator hierarchy:} a comparative study of the Argonne AV family (AV4$'$, AV6$'$, AV8$'$, AV18) demonstrating that the tensor and spin-orbit forces contribute dominantly to the emergence of the Bell-triplet window. 
			\item \textbf{Teleportation preliminaries:} a brief, self-contained review of the standard qubit-teleportation protocol (notation and key steps) to establish the framework used in the main text.
			\item \textbf{Proton spin teleportation at $(151~\mathrm{MeV},90^\circ)$:} a step-by-step derivation showing that the spin amplitude acts as a Bell-state transition,
			$M \propto \ket{\Psi^{+}}\!\bra{\Phi^{-}}$.
			\item \textbf{Receiver-proton spin readout:} a secondary-scattering polarimetry estimate for the outgoing receiver proton, including the analyzing power, useful efficiency, figure of merit, and expected statistical precision.
			\item \textbf{Low-energy proton spin teleportation ($E<10~\mathrm{MeV}$):} a complementary derivation for the $S$-wave-dominated regime where the spin amplitude reduces to a Bell-singlet projector,
			$M \propto \ket{\Psi^{-}}\!\bra{\Psi^{-}}$.
			\item \textbf{Supplemental figures:} additional plots supporting the feasibility analysis and the teleportation derivations.
		\end{itemize}
	\end{abstract}
	
	\maketitle
\end{CJK*}

\clearpage

\setcounter{figure}{0}
\setcounter{table}{0}
\renewcommand{\thefigure}{S\arabic{figure}}
\renewcommand{\thetable}{S\arabic{table}}
\renewcommand{\theequation}{S\arabic{equation}}
\renewcommand{\theHfigure}{Extended Data Fig. \thefigure}
\renewcommand{\theHtable}{Extended Data Table \thetable}

\section{Triplet sources and nuclear Bell tests}

The identification of the Bell-triplet state at $(E_\odot,\theta_\odot)$ establishes unpolarized $pp$ scattering as a promising source for proton triplet pairs. The feasibility of large-scale production is supported by the sizable differential cross section of 3.72\,mb/sr at this kinematic point. Under typical experimental conditions with a solid-angle acceptance of $\Delta\Omega=0.01\ \mathrm{sr}$, a proton beam current of $100\ \mathrm{nA}$, and a 0.10 cm thick liquid hydrogen target, the estimated production rate of these triplet pairs reaches $\sim 10^5\ \text{s}^{-1}$.  

\begin{figure}[htbp]
	\centering
	\includegraphics[width=1\linewidth]{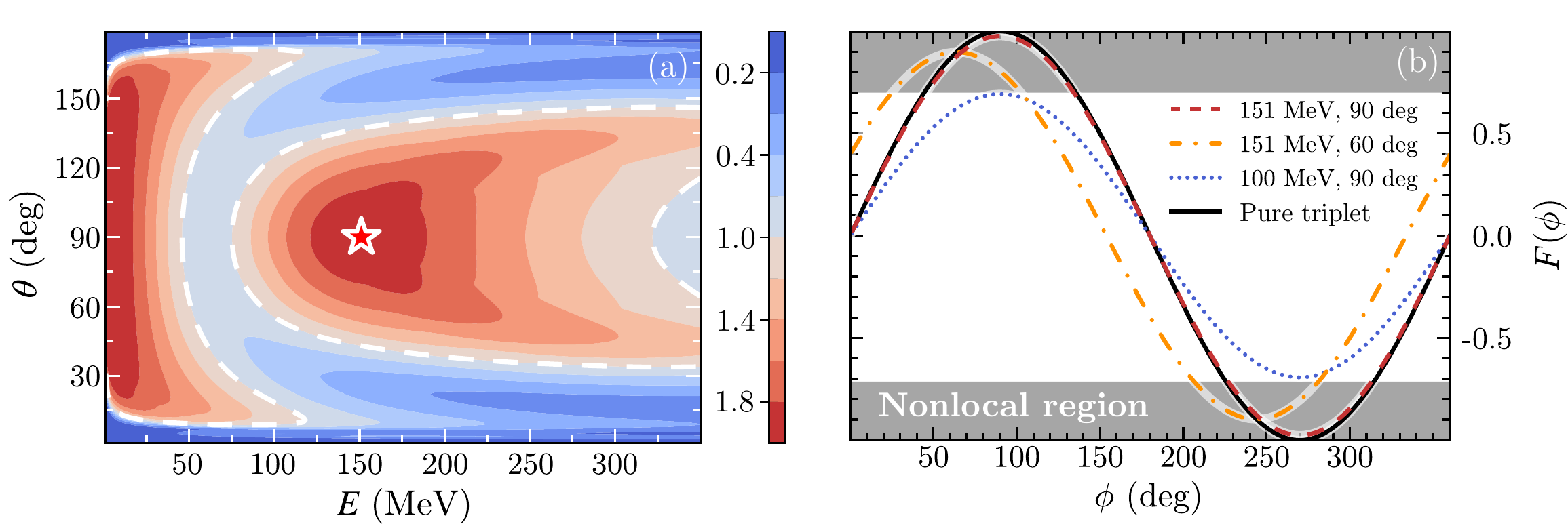}
	\caption{(a) Horodecki function $H(\rho_f)$ derived from the Nijmegen PWA93 database.
		(b) Spin correlation function
		$F(\phi)=\text{Tr}[ (\bm{\sigma}_1\cdot\bm{n}_1) \otimes (\bm{\sigma}_2\cdot\bm{n}_2)\rho_f]$ 
		computed from the same database. The measurement geometry is defined by a fixed analyzer along $\bm{n}_1=(1,0,0)$ and a second analyzer rotating in the $x-z$ plane as $\bm{n}_2=(\sin\phi,0,\cos\phi)$. }
	
	\label{fig:four}
\end{figure}	

The availability of proton triplet pairs opens new avenues for \textit{nuclear Bell tests}, which thus far have relied exclusively on proton singlet pairs.
To assess the feasibility of Bell-inequality violations in the triplet regime, we adopt the Horodecki criterion \cite{Horodecki1995,Brunner2014}, which identifies the kinematic regions where the Clauser-Horne-Shimony-Holt (CHSH) inequality \cite{Clauser1969} can be violated. Figure~\ref{fig:four}(a) shows a contour map of the Horodecki function $H(\rho_f)=u_{1}+u_{2}$, where $u_{1,2}$ are the two largest eigenvalues of the matrix \(U=\tilde{U}^{T}\tilde{U}\), with correlation-tensor elements $\tilde{U}_{ij}=\mathrm{Tr}(\rho_f\sigma_i\otimes\sigma_j)$. A violation of the CHSH inequality is signaled by $H(\rho_f)>1$. The white dashed curve, representing the contour \(H(\rho_f)=1\), separates the CHSH-violating (red) and non-violating (blue) domains.
The pronounced maximum at $(E_\odot,\theta_\odot)$, marked by a red pentagram, together with the surrounding dark-red area, delineates a broad kinematic region suitable for nuclear Bell tests using triplet proton pairs.

The spin correlation function $F(\bm{n}_1,\bm{n}_2) = \text{Tr}[(\bm{\sigma}_1\cdot\bm{n}_1) \otimes (\bm{\sigma}_2\cdot\bm{n}_2)\rho_f]$ 
is commonly measured in experimental tests of Bell inequalities. 
Using the spin density matrix obtained from the PWA93 database, we evaluate $F$ under various conditions, as shown in Fig.\ \ref{fig:four}(b).
In the calculation, the first measurement axis is fixed along the $x$-axis, $\bm{n}_1=(1,0,0)$, while the second axis is rotated by an angle $\phi$ in the $x-z$ plane, $\bm{n}_2=(\sin\phi,0,\cos\phi)$.
For comparison, the ideal pure triplet state 
$\ket{\Psi^+}$ yields $F_{\text{triplet}}(\phi)=\sin\phi$.
The shaded regions correspond to $|F(\phi)| > 1/\sqrt{2}$, the threshold for Bell-inequality violation, signifying that the spin correlations are strong enough to confirm quantum entanglement.
Remarkably, even without further purification, the outgoing proton pairs at $(151\ \text{MeV}, 90^\circ)$ retain pronounced potential for Bell tests, requiring no fine tuning of beam energy or scattering angle.

\textit{Detection-efficiency and locality loopholes}---Here, we briefly discuss the detection efficiency and locality loopholes associated with our triplet nuclear Bell test, assuming an experimental setup similar to that in Ref.\ \cite{LamehiRachti1976}.

(i) \textit{Detection-efficiency loophole}: In this work, carbon polarimetry is used to measure spin correlations of outgoing proton pairs. As a result, a local left-right outcome is obtained only when the proton undergoes a secondary \(pC\) scattering in the analyzer. Events without such a secondary scattering are absent from the reconstructed correlation sample, so the extracted CHSH correlations rely on a fair-sampling assumption for the accepted double-scattering events.

(ii) \textit{Locality loophole}: In the proposed setup, the spin-analysis directions are inferred from the azimuthal bins of passive \(pC\) scattering, rather than chosen actively and independently on an event-by-event basis after the pair is produced. A strict locality test would require
\(L_{AB}>c\tau_{\rm loc}\),
where \(L_{AB}\) is the distance between the two analyzer stations and \(\tau_{\rm loc}\) is the relevant local time interval from an active basis choice to irreversible outcome registration. Such a configuration lies beyond the correlation-measurement setup considered here.
Therefore, although \(pp\) scattering near \(151~{\rm MeV}\) and \(90^\circ\) produces a high-fidelity Bell-triplet spin state that satisfies the Horodecki criterion and is therefore predicted to violate the CHSH inequality for suitably chosen local projective spin-measurement directions, the proposed first-generation setup is not sufficient to realize a loophole-free Bell test.

\newpage

\section{Realistic interactions and the emergent Bell-triplet window}
\label{sec:AVfamily}

To test the robustness and microscopic origin of the emergent Bell-triplet window beyond $\chi$EFT, we repeat the entanglement analysis with the Argonne AV family~\cite{Wiringa1995,Wiringa2002}, whose operator content can be truncated in a controlled manner. Specifically, as summarized in Table~\ref{tab:AVfamily_W10}, AV4$^\prime$ retains only central spin-isospin operators (no tensor and no spin-orbit), AV6$^\prime$ adds the tensor operator but still omits spin-orbit terms, AV8$^\prime$ further includes spin-orbit operators, and AV18 contains the full operator structure.

\begin{table}[htbp]
	\caption{Bell-triplet weight $W_{10}$ at $(E,\theta)=(151~\mathrm{MeV},90^\circ)$ and its maximum over the scanned kinematics for Argonne interactions with controlled operator content.}
	\label{tab:AVfamily_W10}
	\begin{ruledtabular}
		\begin{tabular}{l l c c c c}
			Potential & Operators & Tensor & Spin-orbit & $W_{10}(151~\mathrm{MeV},90^\circ)$ & $\max W_{10}$ \\
			\hline
			AV18         & 18 (full) & Yes & Yes & 0.9885 & 0.9889 \\
			AV8$^\prime$ & 8 (central + $\bm{\sigma}\!\cdot\!\bm{\sigma}$ + tensor + spin-orbit) & Yes & Yes & 0.9882 & 0.9927 \\
			AV6$^\prime$ & 6 (central + $\bm{\sigma}\!\cdot\!\bm{\sigma}$ + tensor)      & Yes & No  & 0.4633 & 0.4633 \\
			AV4$^\prime$ & 4 (central + $\bm{\sigma}\!\cdot\!\bm{\sigma}$)               & No  & No  & 0.0001 & 0.3331 \\
		\end{tabular}
	\end{ruledtabular}
\end{table}

\begin{figure}[htbp]
	\centering
	\includegraphics[width=0.98\textwidth]{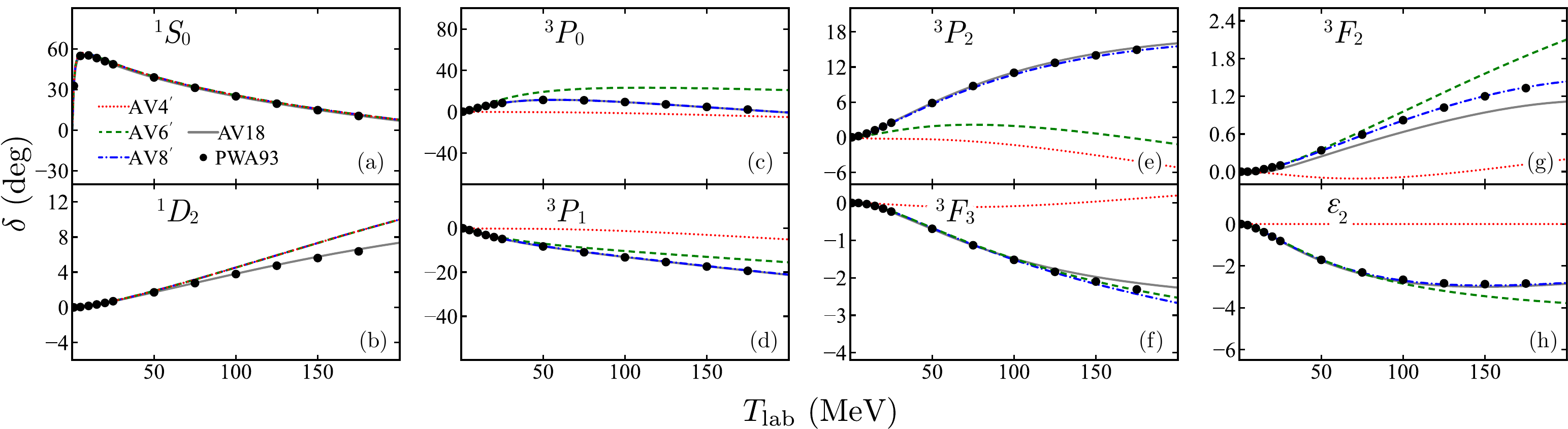}
	\caption{Representative $pp$ phase shifts obtained with AV4$^\prime$ (dotted), AV6$^\prime$ (dashed), AV8$^\prime$ (dash-dotted), and AV18 (solid), compared with the Nijmegen PWA93 analysis (symbols).}
	\label{fig:phaseshifts_AVfamily}
\end{figure}

\begin{figure}[htbp]
	\centering
	\includegraphics[width=0.6\textwidth]{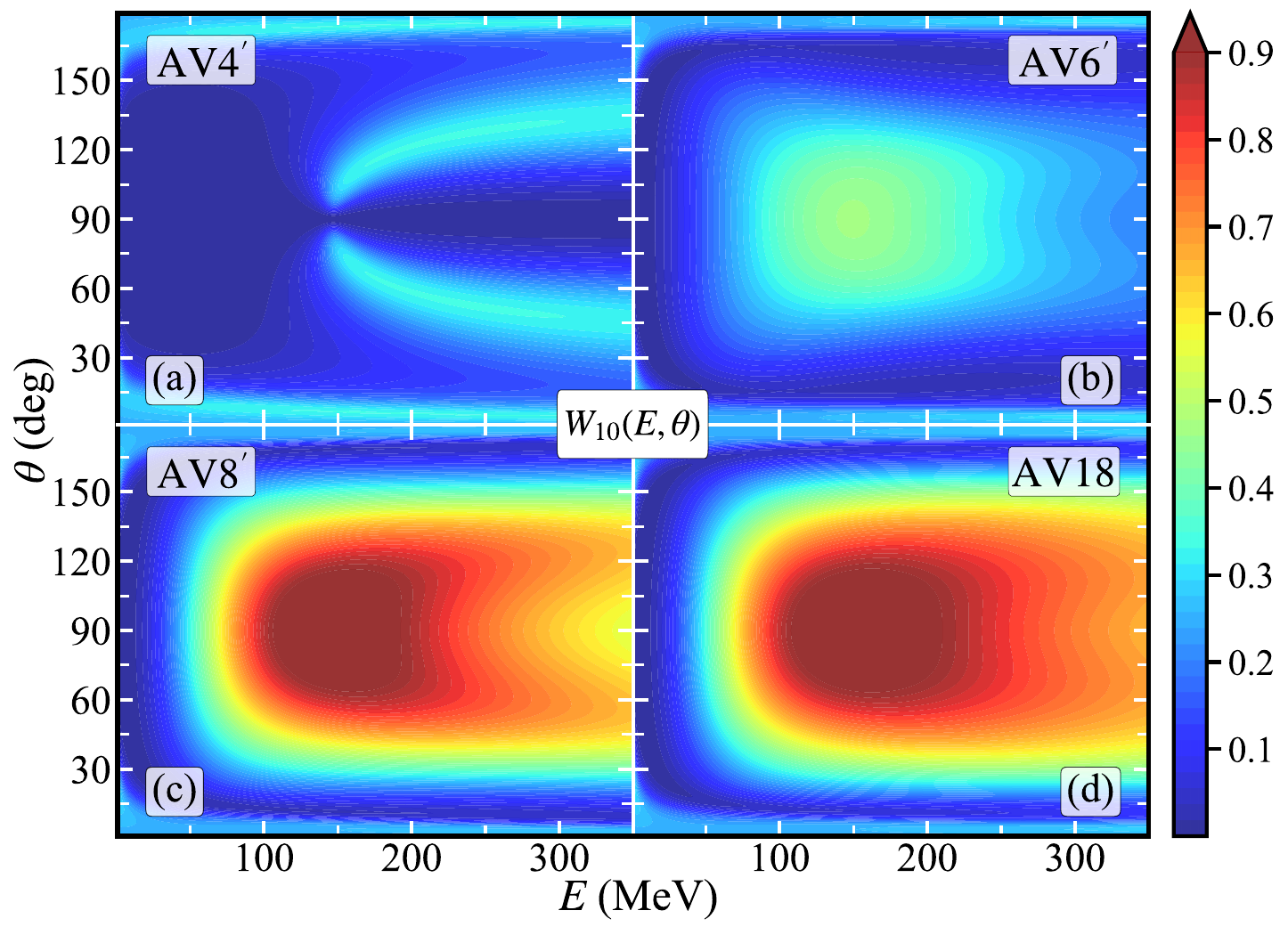}
	\caption{Bell-triplet weight $W_{10}(E,\theta)$ computed with Argonne interactions.
		The entangled triplet window around $(E_\odot,\theta_\odot)$ is reproduced by interactions with a strong tensor sector (AV8$^\prime$, AV18),
		while it is absent without tensor terms (AV4$^\prime$) and strongly suppressed in truncated cases.}
	\label{fig:W10_AVfamily}
\end{figure}

Although these interactions roughly reproduce the dominant low-energy trends of the $pp$ phase shifts (Fig.~\ref{fig:phaseshifts_AVfamily}), their entanglement responses differ qualitatively. Figure~\ref{fig:W10_AVfamily} shows that the near-pure Bell-triplet window around $(E_\odot,\theta_\odot)$ appears only when a sufficiently strong tensor sector is present, and it disappears when tensor terms are removed (AV4$^\prime$).

The contrast between AV6$^\prime$ and AV8$^\prime$ isolates the impact of spin-orbit dynamics once the tensor force is present (see Fig.~\ref{fig:W10_AVfamily} and Table~\ref{tab:AVfamily_W10}). With tensor but no spin-orbit, AV6$^\prime$ already produces a non-negligible population of the $m_s=0$ triplet sector, leading to a localized enhancement of $W_{10}(E,\theta)$ around $(150~\mathrm{MeV},90^\circ)$ in Fig.~\ref{fig:W10_AVfamily}. However, the enhancement remains moderate ($W_{10}\simeq 0.46$ at the benchmark point) and does not develop into a near-unity window. When realistic spin-orbit operators are included (AV8$^\prime$), the effect is amplified dramatically: $W_{10}$ rises from $\sim 0.46$ to $\sim 0.99$ at $(151~\mathrm{MeV},90^\circ)$. (For the $\chi$EFT interaction considered in the main text, the corresponding spin-orbit enhancement is typically subleading relative to the tensor effect.) Microscopically, the tensor interaction provides the essential magnetic-substate mixing that feeds the $m_s=0$ triplet channel, while the spin-orbit interaction reshapes the interference among the dominant partial waves in this energy range (notably the $P$ waves), suppressing competing contributions and allowing the tensor-driven Bell-triplet channel to dominate.

A comparison between AV8$^\prime$ and AV18 further shows that the additional operator structures in AV18 (e.g., quadratic spin-orbit and charge-dependent terms) have only a minor impact on the Bell-triplet window and on $W_{10}$ at the benchmark point: the window is essentially saturated once tensor and spin-orbit operators are included. Conversely, AV4$^\prime$ demonstrates that tensor dynamics are indispensable. Without the tensor operator $S_{12}$, $W_{10}$ is essentially zero at $\theta=90^\circ$ and never becomes large anywhere in the scan (Table~\ref{tab:AVfamily_W10}), consistent with the absence of tensor-induced spin mixing.

Finally, at low energies ($E\lesssim 50$~MeV) all four potentials yield $W_{10}\approx 0$, reflecting the familiar ${}^1S_0$ singlet dominance in the $pp$ channel enforced by the Pauli principle and $S$-wave scattering. Taken together, Table~\ref{tab:AVfamily_W10} and Figs.~\ref{fig:W10_AVfamily} and \ref{fig:phaseshifts_AVfamily} establish a clear hierarchy for the dynamical origin of the Bell-triplet window: the tensor force provides the essential seed, the spin-orbit interaction acts as a powerful amplifier in the relevant kinematic window, and the remaining higher-order operator structures in AV18 play only a subleading role for $W_{10}$.

\newpage

\section{The standard quantum teleportation}

Quantum teleportation is a fundamental protocol in quantum information science that enables the transfer of an unknown quantum state between distant parties using shared entanglement and classical communication \cite{Bennett1993b}. Here we provide a self-contained introduction to the standard protocol, which contributes to the theoretical basis for the proton spin teleportation scheme discussed in the main text.

Consider two parties, Alice and Bob, who wish to transfer an unknown qubit state from Alice to Bob. The protocol involves three qubits: qubit 1 (the data qubit held by Alice carrying the unknown state), and qubits 2 and 3 (an entangled pair shared between Alice and Bob, with Alice holding qubit 2 and Bob holding qubit 3).

The unknown state to be teleported is
\begin{equation}
	\ket{\psi}_1 = \alpha\ket{0}_1 + \beta\ket{1}_1,
	\label{eq:unknown}
\end{equation}
where $\alpha$ and $\beta$ are complex coefficients satisfying $|\alpha|^2+|\beta|^2=1$. Alice and Bob share a maximally entangled Bell state, taken here to be
\begin{equation}
	\ket{\Psi^-}_{23} = \frac{1}{\sqrt{2}}\left(\ket{01}_{23} - \ket{10}_{23}\right).
	\label{eq:epr}
\end{equation}
The four Bell states, which form an orthonormal basis for the two-qubit Hilbert space, are given by
\begin{align}
	\ket{\Psi^\pm}_{23} &= \frac{1}{\sqrt{2}}\left(\ket{01}_{23} \pm \ket{10}_{23}\right), \nonumber\\
	\ket{\Phi^\pm}_{23} &= \frac{1}{\sqrt{2}}\left(\ket{00}_{23} \pm \ket{11}_{23}\right).
	\label{eq:bellbasis}
\end{align}

The initial state of the three-qubit system is the product state $\ket{\psi}_1 \otimes \ket{\Psi^-}_{23}$. Expanding this in terms of the Bell basis for qubits 1 and 2 yields
\begin{align}
	\ket{\psi}_1 \otimes \ket{\Psi^-}_{23}
	&= \left(\alpha\ket{0}_1 + \beta\ket{1}_1\right) \otimes \frac{1}{\sqrt{2}}\left(\ket{01}_{23} - \ket{10}_{23}\right) \nonumber\\
	&= \frac{1}{2}\Big[\ket{\Phi^+}_{12} \otimes \left(\alpha\ket{1}_3 - \beta\ket{0}_3\right) + \ket{\Phi^-}_{12} \otimes \left(\alpha\ket{1}_3 + \beta\ket{0}_3\right) \nonumber\\
	&\quad + \ket{\Psi^+}_{12} \otimes \left(-\alpha\ket{0}_3 + \beta\ket{1}_3\right) + \ket{\Psi^-}_{12} \otimes \left(-\alpha\ket{0}_3 - \beta\ket{1}_3\right)\Big].
	\label{eq:expansion}
\end{align}

Alice then performs a Bell measurement on qubits 1 and 2, projecting them onto one of the four Bell states with equal probability of $1/4$. 

In a gate-based realization, a Bell measurement is implemented by first applying a Bell-state decoder to qubits 1 and 2 and then measuring them in the computational basis. It is useful to define the corresponding Bell-state encoder,
\begin{equation}
	U_{\rm enc}
	=
	\mathrm{CNOT}_{1\rightarrow 2}
	(H_1\otimes \mathbf{1}_2),
\end{equation}
where \(H_1\) denotes the Hadamard gate acting on qubit 1, and \(\mathrm{CNOT}_{1\rightarrow 2}\) is the controlled-NOT gate with qubit 1 as the control and qubit 2 as the target.
This encoder maps the computational product basis to the Bell basis.

The Bell-state decoder is the inverse operation,
\begin{equation}
	U_{\rm dec}
	\equiv
	U_{\rm enc}^{\dagger}
	=
	(H_1\otimes \mathbf{1}_2)\mathrm{CNOT}_{1\rightarrow 2}.
\end{equation}
Its action on the four Bell states is
\begin{equation}
	\begin{aligned}
		U_{\rm dec}|\Phi^+\rangle_{12} &= |00\rangle_{12},\qquad
		U_{\rm dec}|\Psi^+\rangle_{12} = |01\rangle_{12},\\
		U_{\rm dec}|\Phi^-\rangle_{12} &= |10\rangle_{12},\qquad
		U_{\rm dec}|\Psi^-\rangle_{12} = |11\rangle_{12}.
	\end{aligned}
\end{equation}
Applying \(U_{\rm dec}\otimes \mathbf{1}_3\) to the state in Eq.~\eqref{eq:expansion}, we obtain
\begin{equation}
	\begin{aligned}
		(U_{\rm dec}\otimes \mathbf{1}_3)\ket{\psi}_1 \otimes \ket{\Psi^-}_{23}
		=
		\frac{1}{2}\Big[
		&|00\rangle_{12}\otimes(\alpha |1\rangle_3-\beta |0\rangle_3)
		+|10\rangle_{12}\otimes(\alpha |1\rangle_3+\beta |0\rangle_3)\\
		&+|01\rangle_{12}\otimes(-\alpha |0\rangle_3+\beta |1\rangle_3)
		+|11\rangle_{12}\otimes(-\alpha |0\rangle_3-\beta |1\rangle_3)
		\Big].
	\end{aligned}
\end{equation}
Qubits 1 and 2 can then be measured in the computational basis. The four
possible outcomes \(|00\rangle_{12}\), \(|10\rangle_{12}\), \(|01\rangle_{12}\),
and \(|11\rangle_{12}\) identify, respectively, the four Bell components
\(|\Phi^+\rangle_{12}\), \(|\Phi^-\rangle_{12}\),
\(|\Psi^+\rangle_{12}\), and \(|\Psi^-\rangle_{12}\) present in Eq.~\eqref{eq:expansion}.

The measurement outcome, encoded by two classical bits, is transmitted to Bob. Depending on the result, Bob's qubit 3 is left in one of four states, each related to the original state $\ket{\psi}$ by a known unitary transformation:
\begin{align}
	\ket{\Phi^+}_{12}: &\quad \ket{\psi'}_3 = \alpha\ket{1}_3 - \beta\ket{0}_3, \nonumber\\
	\ket{\Phi^-}_{12}: &\quad \ket{\psi'}_3 = \alpha\ket{1}_3 + \beta\ket{0}_3, \nonumber\\
	\ket{\Psi^+}_{12}: &\quad \ket{\psi'}_3 = -\alpha\ket{0}_3 + \beta\ket{1}_3, \nonumber\\
	\ket{\Psi^-}_{12}: &\quad \ket{\psi'}_3 = -\alpha\ket{0}_3 - \beta\ket{1}_3.
	\label{eq:outcomes}
\end{align}

Upon receiving Alice's two classical bits, Bob applies the appropriate Pauli correction to recover the original state:
\begin{align}
	\ket{\Phi^+}_{12} &\Rightarrow ZX\ket{\psi'}_3 = \alpha\ket{0}_3 + \beta\ket{1}_3, \nonumber\\
	\ket{\Phi^-}_{12} &\Rightarrow X\ket{\psi'}_3 = \alpha\ket{0}_3 + \beta\ket{1}_3, \nonumber\\
	\ket{\Psi^+}_{12} &\Rightarrow Z\ket{\psi'}_3 = \alpha\ket{0}_3 + \beta\ket{1}_3, \nonumber\\
	\ket{\Psi^-}_{12} &\Rightarrow \mathbf{1}\ket{\psi'}_3 = \alpha\ket{0}_3 + \beta\ket{1}_3,
	\label{eq:recovery}
\end{align}
where $X=\sigma_x$ and $Z=\sigma_z$ are the Pauli matrices. In all cases, the unknown quantum state $\ket{\psi}$ is faithfully transferred to Bob's qubit, completing the teleportation.

\newpage

\section{\texorpdfstring{Proton spin teleportation at $\bm{(E_\odot,\theta_\odot)=(151\ \text{MeV},90^\circ)}$}{S3. Proton spin teleportation at (E,theta)=(151 MeV, 90 degrees)}}

In this section, we provide a detailed step-by-step analysis of the proton spin teleportation protocol at the kinematic point $(E_\odot,\theta_\odot)=(151~\text{MeV},90^\circ)$, where the spin amplitude acts as a Bell transition operator $M\propto\ket{\Psi^+}\bra{\Phi^-}$.

\textit{Setup}---The protocol involves three \emph{spatially separated single-proton modes}, whose spin degrees of freedom serve as qubits. We label these modes by the same indices/colors as in Fig.~\ref{fig:qt}(b):
\begin{itemize}[leftmargin=*]
	\item Mode 1 (gray): the polarized target proton carrying the unknown spin state $\ket{\psi}_1 = \alpha\ket{0}_1 + \beta\ket{1}_1$ to be teleported, where $|\alpha|^2+|\beta|^2=1$.
	\item Mode 2 (red): one proton from an entangled pair, which will scatter with the target.
	\item Mode 3 (blue): the remote partner of the entangled pair, which will receive the teleported state.
\end{itemize}
Although all three particles are protons, the indices $1,2,3$ refer to these modes (paths/arms) and are therefore operationally distinguishable. In the postselected subspace with one proton in each mode, the spin Hilbert space is equivalent to that of three distinguishable qubits; exchange antisymmetry is already encoded in the $pp$ amplitude used to construct $M$.
The entangled proton pair (protons 2 and 3) is prepared in the Bell-singlet state $\ket{\Psi^-}_{23}$ via, e.g., the $p(d,{}^2\text{He})n$ charge-exchange reaction, following the technique of Ref.~\cite{Sakai2006}. The projectile deuteron energy is chosen such that the outgoing protons have a kinetic energy of approximately 151 MeV.

\textit{Spin state at $t_1$}---At time $t_1$, after the entangled pair is produced and before the scattering with the polarized target, the three-proton spin state is the product
\begin{equation}
	\ket{\Psi(t_1)} = \ket{\psi}_1 \otimes \ket{\Psi^-}_{23} = \left(\alpha\ket{0}_1 + \beta\ket{1}_1\right) \otimes \frac{1}{\sqrt{2}}\left(\ket{01}_{23} - \ket{10}_{23}\right).
\end{equation}
To analyze the effect of the subsequent scattering, we expand this state in terms of the Bell basis for qubits 1 and 2:
\begin{align}
	\ket{\Psi(t_1)} &= \frac{1}{2}\Big[\ket{\Phi^+}_{12} \otimes \left(\alpha\ket{1}_3 - \beta\ket{0}_3\right) + \ket{\Phi^-}_{12} \otimes \left(\alpha\ket{1}_3 + \beta\ket{0}_3\right) \nonumber\\
	&\quad + \ket{\Psi^+}_{12} \otimes \left(-\alpha\ket{0}_3 + \beta\ket{1}_3\right) + \ket{\Psi^-}_{12} \otimes \left(-\alpha\ket{0}_3 - \beta\ket{1}_3\right)\Big].
	\label{eq:qs_t1}
\end{align}
This expansion reveals that each Bell-state component for qubits 1 and 2 is correlated with a distinct state of qubit 3.

\textit{Spin state at $t_2$}---At time $t_2$, the proton in mode 2 (red) scatters with the proton in mode 1 (gray target) at 151~MeV with a center-of-mass scattering angle of $90^\circ$. We postselect coincidence events in which the two outgoing protons are registered in two distinct detector arms (centered at $\pm45^\circ$ in the laboratory frame). At this kinematic point, the scattering matrix is well approximated by
\begin{equation}
	M_{12} \approx \lambda\ket{\Psi^+}_{12}\bra{\Phi^-}_{12},
\end{equation}
where $\lambda = -3.845 - i0.058$ is the dominant matrix element. This operator selects the $\ket{\Phi^-}_{12}$ component from the input state and converts it to $\ket{\Psi^+}_{12}$.

Applying $(M_{12}\otimes \mathbf{1}_3)$ to the state in Eq.~(\ref{eq:qs_t1}), we obtain
\begin{align}
	(M_{12}\otimes \mathbf{1}_3)\ket{\Psi(t_1)} &\approx \frac{\lambda}{2}\ket{\Psi^+}_{12}\bra{\Phi^-}_{12}\Big[\ket{\Phi^-}_{12} \otimes \left(\alpha\ket{1}_3 + \beta\ket{0}_3\right) + \cdots\Big] \nonumber\\
	&= \frac{\lambda}{2}\ket{\Psi^+}_{12} \otimes \left(\alpha\ket{1}_3 + \beta\ket{0}_3\right).
\end{align}
After normalization, the post-scattering state becomes
\begin{equation}
	\ket{\Psi(t_2)} = \ket{\Psi^+}_{12} \otimes \left(\alpha\ket{1}_3 + \beta\ket{0}_3\right).
\end{equation}

\textit{Spin state at $t_3$}---At time $t_3$, after the scattering event, the remote proton 3 (blue) carries the spin state
\begin{equation}
	\ket{\tilde{\psi}}_3 = \alpha\ket{1}_3 + \beta\ket{0}_3.
\end{equation}
This state is related to the original unknown state $\ket{\psi}_1 = \alpha\ket{0}_1 + \beta\ket{1}_1$ by an interchange of $\ket{0}\leftrightarrow\ket{1}$, which corresponds to an inversion of the quantization axis. For the purpose of verifying teleportation, one can simply account for this axis inversion when comparing the measured polarization of proton 3 with the known polarization of the original target.

\textit{Receiver-proton spin readout}---The spin state carried by the receiver proton (Mode 3) can be tested with standard secondary-scattering proton polarimetry, the same basic technique used in previous proton-pair Bell-correlation measurements.
Figure~\ref{fig:receiver_spin_readout} shows the corresponding analyzer geometry.

\begin{figure}[H]
	\centering
	\begin{tikzpicture}[>=Stealth, font=\footnotesize, scale=0.9]
		\node[font=\bfseries\normalsize] at (-4.3,3.1) {(a)};
		\draw[->, very thick, black] (-4.3,0) -- (-0.55,0);
		\shade[ball color=protonblue] (-3.3,0) circle (0.23);
		\node[protonblue, below] at (-2.5,-0.18) {outgoing proton ($\sim$151~MeV)};
		\draw[protonblue, thick] (-3.3,0.72) circle (0.16);
		\fill[protonblue] (-3.3,0.72) circle (0.04);
		\node[protonblue, above] at (-3.3,0.86) {proton spin $s_\perp$};
		\shade[left color=gray!30, right color=gray!60] (-0.35,-0.9) rectangle (0.35,0.9);
		\draw[thick] (-0.35,-0.9) rectangle (0.35,0.9);
		\node[below] at (0,-1.0) {spin analyzer};
		\draw[gray!55, dashed] (0,0) -- (4.4,0);
		\node[gray!60!black, right] at (4.4,0) {beam axis};
		\fill[white] (3.1,-2.95) rectangle (3.6,2.95);
		\fill[detfill] (3.1,0.6) rectangle (3.6,2.05);
		\fill[detfill] (3.1,-2.05) rectangle (3.6,-0.6);
		\draw[thick] (3.1,-2.95) rectangle (3.6,2.95);
		\draw[thick] (3.1,2.05) -- (3.6,2.05);
		\draw[thick] (3.1,0.6) -- (3.6,0.6);
		\draw[thick] (3.1,-0.6) -- (3.6,-0.6);
		\draw[thick] (3.1,-2.05) -- (3.6,-2.05);
		\node[font=\large] at (3.35,1.32) {L};
		\node[font=\large] at (3.35,-1.32) {R};
		\draw[->, very thick, black] (0,0) -- (3.1,1.325);
		\node[black] at (2.8,1.55) {$N_L$};
		\draw[->, very thick, black, dashed] (0,0) -- (3.1,-1.325);
		\node[black] at (2.8,-1.55) {$N_R$};
		\draw[thick, densely dotted] (0,0) -- (3.1,0.6);
		\draw[thick, densely dotted] (0,0) -- (3.1,2.05);
		\draw[->] ([shift=(0:0.8)]0,0) arc (0:11:0.8);
		\node at (2.0,0.2) {$\theta_{\min}$};
		\draw[->] ([shift=(0:1.05)]0,0) arc (0:33.5:1.05);
		\node at (1.7,1.45) {$\theta_{\max}$};
		\begin{scope}[shift={(8.7,0)}]
			\node[font=\bfseries\normalsize] at (-2.2,3.1) {(b)};
			\draw[thick] (-2.2,-2.2) rectangle (2.2,2.2);
			\draw[thin] (-2.0,0) -- (2.0,0);
			\draw[->, thin] (0,-2.0) -- (0,1.6);
			\node[above] at (0,1.6) {$\bm n$};
			\draw[very thick, fill=detfill] (108:0.45) arc[start angle=108, end angle=252, radius=0.45]
			-- (252:1.5) arc[start angle=252, end angle=108, radius=1.5] -- cycle;
			\node[font=\large] at (-0.98,-0.4) {L};
			\draw[very thick, fill=detfill] (-72:0.45) arc[start angle=-72, end angle=72, radius=0.45]
			-- (72:1.5) arc[start angle=72, end angle=-72, radius=1.5] -- cycle;
			\node[font=\large] at (0.98,-0.4) {R};
			\draw[<->] (108:0.78) arc[start angle=108, end angle=252, radius=0.78];
			\node[fill=detfill, inner sep=0.5pt] at (-0.78,0) {$2\phi$};
			\draw[<->] (-72:0.78) arc[start angle=-72, end angle=72, radius=0.78];
			\node[fill=detfill, inner sep=0.5pt] at (0.78,0) {$2\phi$};
		\end{scope}
	\end{tikzpicture}
	\caption{Receiver-proton spin readout. (a)~The outgoing receiver proton scatters from a carbon analyzer. The spin-orbit interaction in secondary \(pC\) scattering produces a left-right asymmetry, so that the counts \(N_L\) and \(N_R\) determine the transverse polarization component through the calibrated analyzing power. (b)~Transverse view of the left-right analyzer sectors, each subtending an angle \(2\phi\); \(\bm n\) is the analyzing direction. The reconstructed receiver polarization is compared with the known input polarization, after accounting for the fixed axis inversion in the \(151~\mathrm{MeV}\) protocol.}
	\label{fig:receiver_spin_readout}
\end{figure}

The \(\sim151~\mathrm{MeV}\) receiver proton is sent through the carbon analyzer in Fig.~\ref{fig:receiver_spin_readout}, where the spin-orbit part of the proton-carbon interaction produces a left-right azimuthal asymmetry
\begin{equation}
	\varepsilon=\frac{N_L-N_R}{N_L+N_R}=A_{\rm eff}\,\mathcal{P}_{n},
\end{equation}
with \(A_{\rm eff}\) the acceptance-averaged analyzing power and \(\mathcal{P}_{n}\) the polarization component normal to the analyzer plane.
Rotating the analyzer azimuth, or equivalently combining orthogonal analyzer sectors, gives access to the transverse spin components needed to compare the measured receiver polarization with the known input target polarization after the fixed axis inversion discussed above.

The statistical quality of this readout is governed by the polarimeter figure of merit
\begin{equation}
	\mathrm{FOM}=A_{\rm eff}^{2}\eta,
	\qquad
	\delta\mathcal{P}_{n}=\frac{1}{\sqrt{N\,\mathrm{FOM}}},
\end{equation}
where \(\eta\) is the fraction of incident receiver protons that scatter into the useful left-right acceptance and \(N\) is the number of protons entering the analyzer.
At the relevant energy, proton-carbon data and focal-plane-polarimeter calibrations give \(A_{\rm eff}\simeq0.5\) for useful angular acceptances~\cite{Comparat1974,McNaughton1985,Meyer1983}.
For a representative graphite analyzer of areal density \(2~\mathrm{g/cm^2}\), the useful analyzed fraction is \(\eta\sim\) a few \(\times10^{-3}\), corresponding to \(\mathrm{FOM}\sim10^{-3}\).
Combining this with the teleportation event-rate estimate below, \(\dot N_{\rm tel}\sim10^{3}\)--\(10^{4}~\mathrm{s}^{-1}\), gives a statistical precision of order \(\delta\mathcal{P}_{n}\simeq0.02\) on a timescale ranging from minutes to a few hours, depending on the accepted source rate and analyzer thickness.
The finite analyzing power therefore reduces the statistical efficiency but does not constitute a qualitatively new obstacle to verifying the transferred spin state.

\textit{Comparison with standard teleportation}---In the standard quantum teleportation protocol \cite{Bennett1993b}, Alice first performs a Bell-state decoding operation on qubits 1 and 2 in Eq.~\eqref{eq:expansion}, transforming the Bell basis into the computational basis \cite{Nielsen_Chuang_2010}. She then measures the two qubits and communicates the classical results to Bob, who applies the corresponding Pauli correction ($\mathbf{1}$, $X$, $Z$, or $XZ$). Consequently, as evident from Eq.~\eqref{eq:expansion}, each Bell state of the first two qubits is perfectly correlated with a distinct conditional state of qubit 3.

The proton-proton scattering operation is structurally different from a gate-based Bell measurement. The spin amplitude \(\propto|\Psi^+\rangle_{12}\langle\Phi^-|_{12}\) is not a unitary transformation between the computational product basis and the Bell basis on the full two-qubit Hilbert space, but a Bell transition operator selected by the scattering kinematics. 
Under the prescribed postselection, the spin-dependent \(pp\) amplitude maps, to a very good approximation, the \(|\Phi^-\rangle_{12}\) component of Eq.~\eqref{eq:qs_t1} into the outgoing \(|\Psi^+\rangle_{12}\) channel. The nuclear process therefore realizes a single postselected Bell transition branch, rather than a complete four-outcome Bell-state measurement.

This single Bell-state transition branch is the central implementation-level distinction of the present protocol. In gate-controllable platforms, a Bell measurement can be implemented through well-controlled single- and two-qubit gates followed by projective readout \cite{Riebe2004,Barrett2004,Steffen2013,Nolleke2013}. Such gate-level control is not available for individual hadronic spin qubits in a MeV proton-scattering experiment. Instead, the spin-dependent \(pp\) amplitude and the selected final-state kinematics provide one postselected Bell-state transition branch. Consequently, only the heralding of a successful coincidence event must be communicated; no two-bit outcome-dependent choice among four Pauli corrections is required. The protocol thus trades deterministic gate programmability for kinematic selectivity and realizes a probabilistic, postselected form of teleportation-like spin-state transfer in a strongly interacting massive-particle platform.

\textit{Count-rate estimate}---We estimate the spin-teleportation event rate under baseline experimental conditions.
The entangled pair (protons 2 and 3) is assumed to be produced via the charge-exchange reaction
$p(d,{}^{2}\mathrm{He})n$ and delivered to a polarized proton target (proton 1).
A successful event is tagged by coincident detection of the two scattered protons (1 and 2) at $\pm45^\circ$ in the laboratory frame within an angular window $\Delta\theta=\pm1^\circ$, corresponding to
$\Delta\Omega = 2\pi[\cos(44^\circ)-\cos(46^\circ)] \simeq 0.155~\mathrm{sr}$.

At $151~\mathrm{MeV}$ and $45^\circ_{\rm lab}$ we take $(d\sigma/d\Omega)_{\rm lab}\simeq 10.6~\mathrm{mb/sr}$~\cite{NNOnline}, giving $\sigma_{\rm acc}\simeq (d\sigma/d\Omega)_{\rm lab}\Delta\Omega \simeq 1.64~\mathrm{mb}$. For a $1~\mathrm{cm}$ thick liquid-hydrogen target with areal density $N_t\simeq 4.24\times10^{22}~\mathrm{cm^{-2}}$, the scattering (tagging) probability is $P_{\rm scatter}\simeq \sigma_{\rm acc}N_t \simeq 7.0\times10^{-5}$ per incident proton-2 on target-1. For the upstream $p(d,{}^{2}\mathrm{He})n$ source, integrating the $E_x=0$--$1~\mathrm{MeV}$ forward $^1\mathrm{H}(d,2p)n$ spectra of Ref.\ \cite{Kox1993} over a representative $20~\mathrm{msr}$ acceptance gives $\sigma_{d\to{}^{2}\mathrm{He}}\sim(0.2$--$0.3)~\mathrm{mb}$.  For $I_d=100~\mathrm{nA}$ ($\dot N_d=I_d/e\simeq6.2\times10^{11}~\mathrm{s^{-1}}$) and a $10~\mathrm{cm}$ liquid-hydrogen source target ($N_H\simeq4.3\times10^{23}~\mathrm{cm^{-2}}$), this gives $\dot N_{\rm pair}\sim \dot N_d N_H\sigma_{d\to{}^{2}\mathrm{He}}\simeq6\times10^{7}~\mathrm{s^{-1}}$. Taking this as the incident entangled-proton flux, the resulting spin-teleportation event rate is $\dot N_{\rm tel}\simeq \dot N_{\rm pair}P_{\rm scatter}\simeq 4.2\times10^{3}~\mathrm{s^{-1}}$, which is sufficient to accumulate a statistically significant sample for verification at rates of $\sim10^{3}$--$10^{4}~\mathrm{s^{-1}}$. In the realistic experimental analysis, this coincidence yield will be corrected for the target dilution factor, and accidental coincidences will also be subtracted.

\textit{Extension to mixed states}---The analysis above assumes that the target proton carries a pure spin state. Here we generalize the protocol to the case where the target proton is in a mixed state, which is more realistic for experimental implementations.

A general mixed state of the target proton can be written as
\begin{equation}
	\rho_1 = \sum_i p_i \ket{\psi_i}_1\bra{\psi_i}_1,
	\label{eq:mixed_state}
\end{equation}
where $p_i \geq 0$, $\sum_i p_i = 1$, and each $\ket{\psi_i}_1 = \alpha_i\ket{0}_1 + \beta_i\ket{1}_1$ is a normalized pure state with $|\alpha_i|^2 + |\beta_i|^2 = 1$. The initial three-proton density matrix at time $t_1$ is
\begin{equation}
	\rho(t_1) = \rho_1 \otimes \ket{\Psi^-}_{23}\bra{\Psi^-}_{23} = \sum_i p_i \ket{\psi_i}_1\bra{\psi_i}_1 \otimes \ket{\Psi^-}_{23}\bra{\Psi^-}_{23}.
\end{equation}

At time $t_2$, the scattering operator $M_{12} \otimes \mathbf{1}_3$ acts on the system. Since the scattering transformation is linear, we can apply it to each term in the mixture separately. For each pure-state component, following the derivation in the previous subsections, we have
\begin{equation}
	(M_{12} \otimes \mathbf{1}_3)\left(\ket{\psi_i}_1\bra{\psi_i}_1 \otimes \ket{\Psi^-}_{23}\bra{\Psi^-}_{23}\right)(M_{12}^\dagger \otimes \mathbf{1}_3) \propto \ket{\Psi^+}_{12}\bra{\Psi^+}_{12} \otimes \ket{\tilde{\psi}_i}_3\bra{\tilde{\psi}_i}_3,
\end{equation}
where $\ket{\tilde{\psi}_i}_3 = \alpha_i\ket{1}_3 + \beta_i\ket{0}_3$ is the teleported state corresponding to the input $\ket{\psi_i}_1$.

Summing over all components and normalizing, the post-scattering density matrix becomes
\begin{equation}
	\rho(t_2) = \ket{\Psi^+}_{12}\bra{\Psi^+}_{12} \otimes \tilde{\rho}_3,
\end{equation}
where the reduced density matrix of proton 3 is
\begin{equation}
	\tilde{\rho}_3 = \sum_i p_i \ket{\tilde{\psi}_i}_3\bra{\tilde{\psi}_i}_3.
	\label{eq:teleported_mixed}
\end{equation}

To relate $\tilde{\rho}_3$ to the original mixed state $\rho_1$, we note that each teleported pure-state component $\ket{\tilde{\psi}_i}_3 = \alpha_i\ket{1}_3 + \beta_i\ket{0}_3$ differs from the corresponding input state $\ket{\psi_i}_1 = \alpha_i\ket{0}_1 + \beta_i\ket{1}_1$ only by the interchange $\ket{0}\leftrightarrow\ket{1}$, i.e., an inversion of the quantization axis. Since this transformation applies uniformly to all components in the mixture, the teleported mixed state $\tilde{\rho}_3$ is identical to the original mixed state $\rho_1$ up to the same axis inversion.

Therefore, the teleportation protocol faithfully transfers the mixed state from the target proton to the remote proton, up to the inversion of the quantization axis. This result confirms that the nuclear teleportation protocol preserves not only the coherent superposition of pure states but also the classical probabilistic mixture encoded in the density matrix.

\textit{Kinematic windows for the Bell transition operator}---The triplet-based teleportation scheme relies on the fact that, within a finite kinematic window around $(E_\odot,\theta_\odot)$, the dynamics produces an effective Bell transition operator,
\begin{equation}
	\label{eq:gate_triplet}
	G_\odot \equiv \ket{\Psi^+}\bra{\Phi^-}.
\end{equation}
To quantify how closely the physical $pp$ spin amplitude $M(E,\theta)$ approximates the target gate $G_\odot$,
we employ the normalized trace (Hilbert-Schmidt) similarity~\cite{WatrousTQI}
\begin{equation}
	\label{eq:hs_similarity_triplet}
	\mathcal{S}(E,\theta)
	=
	\frac{\big|\mathrm{Tr}\!\left[M^\dagger(E,\theta)\,G_\odot\right]\big|}
	{\sqrt{\mathrm{Tr}\!\left[M^\dagger(E,\theta)\,M(E,\theta)\right]}},
\end{equation}
which is invariant under an overall rescaling $M\!\to\!\alpha M$ and satisfies $0\le \mathcal{S}\le 1$.
Here $M(E,\theta)$ is constructed from the PWA93 phase shifts at beam energy $E$ and center-of-mass scattering angle $\theta$.

\begin{figure*}[t]
	\centering
	\includegraphics[width=\textwidth]{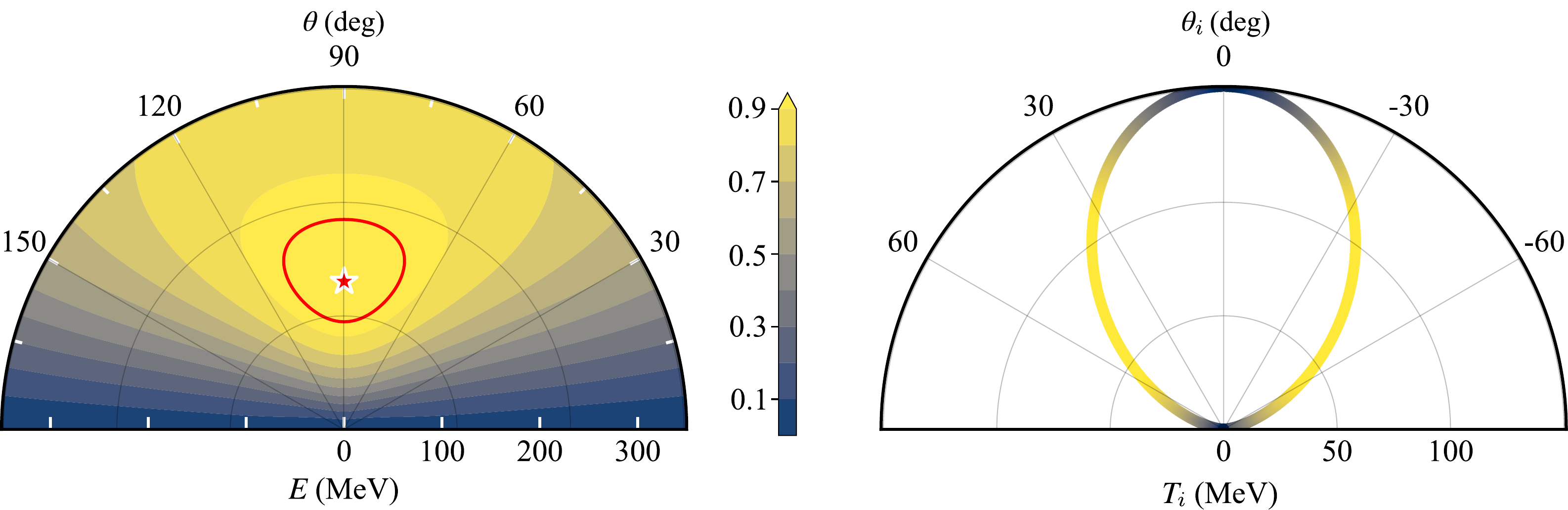}
	\caption{Kinematic windows for the Bell transition operator $\ket{\Psi^+}\bra{\Phi^-}$.
		(a) Similarity $\mathcal{S}(E,\theta)$ between the physical $pp$ spin amplitude $M(E,\theta)$ (from PWA93)
		and $\ket{\Psi^+}\bra{\Phi^-}$ as a function of the beam energy $E$ and center-of-mass angle $\theta$.
		The red contour encloses the region $\mathcal{S}>0.95$ (Bell-triplet window), and the star marks $(E_\odot,\theta_\odot)$.
		(b) Kinematically allowed locus in the $(T_i,\theta_i)$ plane (laboratory kinetic energy and scattering angle) for either outgoing proton $i$ at beam energy $E_\odot$.}
	\label{fig:triplet_windows}
\end{figure*}

Figure~\ref{fig:triplet_windows}(a) displays $\mathcal{S}(E,\theta)$ over a broad energy-angle plane.
A pronounced high-similarity ``island'' forms around $(E_\odot,\theta_\odot)$ (star),
demonstrating that the nuclear interaction naturally generates an amplitude close to $G_\odot$ within a restricted kinematic domain.
We define the \emph{Bell-triplet kinematic window} as the region where $\mathcal{S}(E,\theta)>\mathcal{S}_0$,
with $\mathcal{S}_0=0.95$ (red contour). For clarity, the colormap is saturated at $\mathcal{S}=0.9$.

For experimental guidance, it is useful to translate this window into laboratory-frame observables.
For equal-mass $pp$ scattering in nonrelativistic kinematics, energy-momentum conservation maps a fixed beam energy to a one-dimensional locus in the $(T_i,\theta_i)$ plane (laboratory kinetic energy and scattering angle) for each outgoing proton $i$.
Figure~\ref{fig:triplet_windows}(b) shows this locus for $E_\odot$.
Its overlap with the Bell-triplet window specifies the coincidence-acceptance band for selecting events that implement the Bell-triplet transition with high fidelity.

\newpage

\section{\texorpdfstring{Spin transport and decoherence in proton spin teleportation at $\bm{(E_\odot,\theta_\odot)=(151\ \text{MeV},90^\circ)}$}{Decoherence effect of the Proton spin teleportation}}

In this section, we assess decoherence of the entangled proton pair. The pair is created as a spin singlet in the $p(d,{}^{2}\mathrm{He})n$ charge-exchange reaction using a $\sim305~\mathrm{MeV}$ deuteron beam, with the quoted beam energy accounting for the negative reaction $Q$ value. It then travels through the production target, magnetic elements, vacuum windows, an air gap, and tracking chambers before one proton undergoes the $pp$ scattering that realizes the Bell-state transition. Coherent transport effects can be reconstructed and corrected from the measured momenta and trajectories, whereas unresolved stochastic or irreversible processes cause genuine decoherence. We therefore consider coherent spin precession and event-by-event spin-transport correction, and estimate decoherence from material interactions and ambient-field fluctuations.

We use natural units $c=1$, where $c$ is the speed of light, and a subscript $0$ marks reference kinematics.  For proton rest mass $m_p=938.27~\mathrm{MeV}$, the reference Lorentz factor $\gamma_0$, speed $\beta_0$ in units of $c$, and laboratory momentum $p_0$ are $\gamma_0=1+E_\odot/m_p=1.16$, $\beta_0=({1-\gamma_0^{-2}})^{1/2}=0.51$, and $p_0=m_p\gamma_0\beta_0=553.32~\mathrm{MeV}$.  The proton magnetic anomaly is $G\equiv(g_p-2)/2=1.79$, where $g_p$ is the proton $g$ factor~\cite{Mohr2025}.

\emph{Coherent spin transport}---For a transverse dipole field that bends the reference orbit through an angle $\Theta$, the Thomas-Bargmann-Michel-Telegdi (BMT) and cyclotron frequencies give the laboratory spin-rotation angle $\phi(\Theta)$~\cite{Bargmann1959,Jackson1999}:
\begin{equation}
 \phi(\Theta)=(1+\gamma_0G)\Theta,
 \qquad
 \phi_0\equiv\phi(\Theta_0)
 =(1+\gamma_0G)\frac{\pi}{2}=277.32^\circ ,
 \label{eq:compact_bmt_rotation}
\end{equation}
where $\Theta_0\equiv\pi/2$ is the chosen $90.00^\circ$ orbital bend and $\phi_0$ is its associated spin rotation.  The same rotation applied to both protons is a local unitary and leaves the singlet invariant.  If their trajectories instead generate spin angles $\phi_A$ and $\phi_B$, differences among the accepted trajectories allow the relative angle $\Delta\phi\equiv\phi_A-\phi_B$ to vary from event to event.  For a given $\Delta\phi$, the transported state is
\begin{equation}
 \ket{\Psi^-(\Delta\phi)}
 =\frac{\exp(-\mathrm{i}\Delta\phi/2)\ket{\uparrow\downarrow}
 -\exp(\mathrm{i}\Delta\phi/2)\ket{\downarrow\uparrow}}{\sqrt2},
 \qquad
 \mathcal C=\left|\left\langle
 \exp(\mathrm{i}\Delta\phi)\right\rangle\right|.
 \label{eq:compact_phase_average}
\end{equation}
Here $\ket{\uparrow}$ and $\ket{\downarrow}$ are spin projections along the fixed axis, and each two-arrow ket lists protons $A$ and $B$ in that order.  The ket $\ket{\Psi^-(\Delta\phi)}$ denotes the state for one event.  Because $\Delta\phi$ varies across the accepted sample, $\langle\cdots\rangle$ denotes the accepted-event average of the phase factor $\exp(\mathrm{i}\Delta\phi)$, and $\mathcal C\in[0,1]$ is the concurrence of the resulting ensemble.  Thus each event remains maximally entangled, whereas averaging over their relative phases can reduce $\mathcal C$; for a Gaussian distribution of $\Delta\phi$ with standard deviation $\sigma_\phi$, $\mathcal C=\exp(-\sigma_\phi^2/2)$~\cite{Wootters1998,YuEberly2006}.

The momenta and trajectories can be reconstructed event by event, as demonstrated for the same source reaction by Sakai \emph{et al.}~\cite{Sakai2006}. Their selection $E_{\mathrm{rel}}<1.00~\mathrm{MeV}$, where $E_{\mathrm{rel}}$ is the total kinetic energy of the two protons in their pair center-of-mass frame, limits the magnitude $q$ of either proton's momentum in that frame:
\begin{equation}
 q\equiv\frac12\sqrt{E_{\mathrm{rel}}(E_{\mathrm{rel}}+4m_p)}
 \simeq\sqrt{m_pE_{\mathrm{rel}}},
 \qquad
 p_{A,\parallel}-p_{B,\parallel}
 =2\Gamma_{\mathrm{pair}}q_\parallel .
 \label{eq:Erel_q}
\end{equation}
The symbol $\parallel$ labels the boost direction: $q_\parallel$ is the signed center-of-mass component, $p_{A,\parallel}$ and $p_{B,\parallel}$ are the laboratory components, and $\Gamma_{\mathrm{pair}}\simeq\gamma_0$ is the pair-center-of-mass Lorentz factor. For a deliberately maximal-phase benchmark, we take representative mean energy $\overline E_{\mathrm{rel}}=0.50~\mathrm{MeV}$ and $|q_\parallel|=q$.  Defining the normalized momentum difference $\delta_p\equiv(p_{A,\parallel}-p_{B,\parallel})/p_0$, its root-mean-square (RMS) spread is
$\sigma_{\delta_p}\simeq2\gamma_0 ({m_p\overline E_{\mathrm{rel}}})^{1/2}/p_0=0.09$.

For a fixed magnetic-field integral, let $p$ denote a proton's laboratory momentum and define
\begin{equation}
 K\equiv p_0\left.\frac{d\phi}{dp}\right|_{p_0}
 =-\Theta_0\left(1+\frac{G}{\gamma_0}\right)
 =-4.00~\mathrm{rad},
 \qquad
 \mathcal C_{\mathrm{untracked}}\simeq
 \exp(-K^2\sigma_{\delta_p}^2/2)=0.94 .
 \label{eq:compact_momentum_dephasing}
\end{equation}
Thus $K$ gives the spin-angle change per unit relative momentum deviation, while $\mathcal C_{\mathrm{untracked}}$ is the concurrence without event-by-event correction. Correction instead leaves the residual reconstruction width $\sigma_{\delta_p,\mathrm{res}}$ and concurrence
$\mathcal C_{\mathrm{corr}}$:
\begin{equation}
 \mathcal C_{\mathrm{corr}}
 =\exp\!\left(-\frac{K^2\sigma_{\delta_p,\mathrm{res}}^2}{2}\right).
 \label{eq:corrected_transport_concurrence}
\end{equation}
For illustration, assigning each proton an independent relative momentum error with an RMS value of $10^{-2}$ gives $\sigma_{\delta_p,\mathrm{res}}=\sqrt2\times10^{-2}$ and $1-\mathcal C_{\mathrm{corr}}=1.60\times10^{-3}$.

\emph{Material-induced dephasing}---We use an average $0.050~\mathrm{g\cdot cm^{-2}}$ polyethylene path, two $125.00~\mu\mathrm m$ Kapton windows, $30.00~\mathrm{cm}$ of dry air, and two $120.00~\mu\mathrm m$ Mylar layers.
For material layer $j$, let $x_j$ be its areal mass density and $X_{0,j}$ its radiation length (both in $\mathrm{g\cdot cm^{-2}}$); the sum $x/X_0\equiv\sum_jx_j/X_{0,j}$ is the total material budget in radiation lengths.  Particle Data Group (PDG) inputs and the relativistic Bohr
variance give~\cite{PDG2024,LynchDahl1991,Leo1994}
\begin{align}
 \frac{x}{X_0}&=\sum_j\frac{x_j}{X_{0,j}}=3.82\times10^{-3},\\
 \theta_0&=\frac{13.6~\mathrm{MeV}}{\beta_0p_0}
 \sqrt{\frac{x}{X_0}}
 \left[1+0.038\ln\!\left(\frac{x/X_0}{\beta_0^2}\right)\right]
 =2.51~\mathrm{mrad},\\
 \sigma_E^2&=(0.1569~\mathrm{MeV^2\,cm^2\,g^{-1}})\,\gamma_0^2
 \left(1-\frac{\beta_0^2}{2}\right)
 \sum_j\left\langle\frac{Z}{A}\right\rangle_jx_j,
 \qquad \sigma_E=0.12~\mathrm{MeV}.
 \label{eq:compact_material_inputs}
\end{align}
Here $\theta_0$ is the projected RMS scattering angle of one proton, $\sigma_E$ its RMS energy-loss fluctuation, and $\langle Z/A\rangle_j$ the composition-averaged charge-to-mass ratio of layer $j$; in this ratio $Z$ and $A$ are the atomic number and relative atomic mass,
not proton labels.  For a simple estimate, we place all material before the full bend and treat the two protons independently, giving the RMS relative phase $\sigma_{\phi,\mathrm{mat}}$ and material-limited concurrence $\mathcal C_{\mathrm{mat}}$:
\begin{equation}
 \sigma_{\phi,\mathrm{mat}}
 =\sqrt{2(1+\gamma_0G)^2\theta_0^2
 +2K^2\left(\frac{\sigma_E}{\beta_0p_0}\right)^2}
 =1.12\times10^{-2}~\mathrm{rad},
 \qquad
 1-\mathcal C_{\mathrm{mat}}=6.30\times10^{-5}.
 \label{eq:compact_material_loss}
\end{equation}
The two factor-of-two terms are the independent-proton contributions from multiple scattering and energy straggling.  This geometry proxy is a conditional design estimate; an actual value requires material locations and the spin-transfer map.

\emph{Nuclear reactions}---Let $\lambda_{I,j}$ be the nuclear interaction length of layer $j$ in $\mathrm{g\cdot cm^{-2}}$, with $I$ denoting an inelastic interaction.  For a given proton, the reaction probability $P_R$ ($R$ labels reaction) is
\begin{equation}
 P_R=1-\exp\!\left(-\sum_j\frac{x_j}{\lambda_{I,j}}\right)
\simeq1.85\times10^{-3},
 \label{eq:compact_reaction_probability}
\end{equation}
using tabulated interaction lengths, with the scale near this energy checked against proton-carbon reaction data~\cite{PDG2024,Carlson1996}.  Let $f_{\mathrm{dep}}$ be the fraction of reactions that completely randomize a proton spin.  Let $\eta_R$ parameterize any selection bias between reaction and no-reaction events, with $\eta_R=1$ denoting equal acceptance.  The probability that an accepted proton has undergone complete spin randomization is then
\begin{equation}
 p_{\mathrm{acc}}
 =f_{\mathrm{dep}}
 \frac{\eta_RP_R}{1-P_R+\eta_RP_R},
 \qquad
 \mathcal C_{\mathrm{nuc}}
 =\max\!\left[0,\frac{3(1-p_{\mathrm{acc}})^2-1}{2}\right].
 \label{eq:compact_nuclear_channel}
\end{equation}
Here $p_{\mathrm{acc}}$ is the probability that a proton in a selected event is completely depolarized, and $\mathcal C_{\mathrm{nuc}}$ is the concurrence after such independent channels act on both protons. For $(f_{\mathrm{dep}},\eta_R)=(0.30,0.10)$, $1-\mathcal C_{\mathrm{nuc}}=1.67\times10^{-4}$; setting $\eta_R=1$ gives $1.66\times10^{-3}$.  The extreme $f_{\mathrm{dep}}=\eta_R=1$ choice gives $5.54\times10^{-3}$.  A quantitative estimate requires $f_{\mathrm{dep}}$ and $\eta_R$ for the actual detector configuration and event selection; the values quoted above are illustrative.

\emph{Ambient fields}---Let $i$ label an accepted event, $t$ the elapsed time, and $t_f$ the flight time to the subsequent $pp$ scattering.  The trajectories $\bm r_A^{(i)}(t)$ and $\bm r_B^{(i)}(t)$ are the proton positions, while $B(\bm r)$ is the magnetic-field component about the precession axis at position $\bm r$.  The time-integrated differential field is $\Phi_i\equiv\int_0^{t_f}\{B[\bm r_A^{(i)}(t)]-B[\bm r_B^{(i)}(t)]\}dt$; its standard deviation across events, expressed as a field, is $\sigma_{\delta B}\equiv {\langle(\Phi_i-\langle\Phi_i\rangle)^2\rangle}^{1/2}/t_f$.

At the reference energy, the BMT equation gives the laboratory-frame angular precession coefficients $\gamma_\perp=[(G+\gamma_0^{-1})/(G+1)]\gamma_p=0.95\,\gamma_p$ and $\gamma_\parallel=\gamma_p/\gamma_0=0.86\,\gamma_p$ for magnetic-field components transverse and longitudinal to the proton velocity, respectively~\cite{Bargmann1959,Jackson1999}.  The symbol $\gamma_p$ is the proton angular gyromagnetic ratio, distinct from $\gamma_0$, with $\gamma_p/2\pi=42.58~\mathrm{MHz/T}$; its rest-frame value is therefore a conservative upper coefficient~\cite{Mohr2025}.  Defining the effective inhomogeneous dephasing time as $T_2^*\equiv1/(\gamma_p\sigma_{\delta B})$, the ambient-field concurrence $\mathcal C_{\mathrm{amb}}$ is estimated by $\mathcal C_{\mathrm{amb}}\simeq
\exp[-(t_f/T_2^*)^2/2]$. For the illustrative reference layout, comprising straight sections of $1.30$, $0.95$, and $1.95~\mathrm m$ and two $45^\circ$ bends of radius $1.60~\mathrm m$, the total flight path is $6.71~\mathrm m$, corresponding to $t_f=44.08~\mathrm{ns}$. Gradients of $10.00$--$100.00~\mu\mathrm{T/m}$ across a maximum $0.30~\mathrm m$ separation imply $\sigma_{\delta B}\lesssim3.00$--$30.00~\mu\mathrm T$ and $T_2^*=0.12$--$1.25~\mathrm{ms}$. At the upper endpoint $\sigma_{\delta B}=30.00~\mu\mathrm T$, the numerical estimate is $1-\mathcal C_{\mathrm{amb}}\simeq6.26\times10^{-8}$; at $3.00~\mu\mathrm T$, the corresponding loss is $6.26\times10^{-10}$. 

Taken together, these estimates indicate that transport-induced decoherence remains small. The proton pair is therefore expected to retain high concurrence when it reaches the subsequent $pp$ scattering. Existing proton-pair Bell experiments~\cite{LamehiRachti1976,Sakai2006} also observed strong spin correlations after magnetic transport and secondary scattering, including the same $^{2}$He production method in Ref.~\cite{Sakai2006}, consistent with the preservation of proton-spin correlations over the relevant flight path.

\newpage

\section{\texorpdfstring{Proton spin teleportation at $\bm{E<10}$ MeV}{Proton spin teleportation at E<10 MeV}}

In this section, we present an alternative teleportation protocol that operates at low energies ($E<10$~MeV), where $S$-wave scattering dominates and the spin amplitude acts as a Bell-singlet projection operator $M\propto\ket{\Psi^-}\bra{\Psi^-}$.

\begin{figure}
	\centering
	\includegraphics[width=0.4\linewidth]{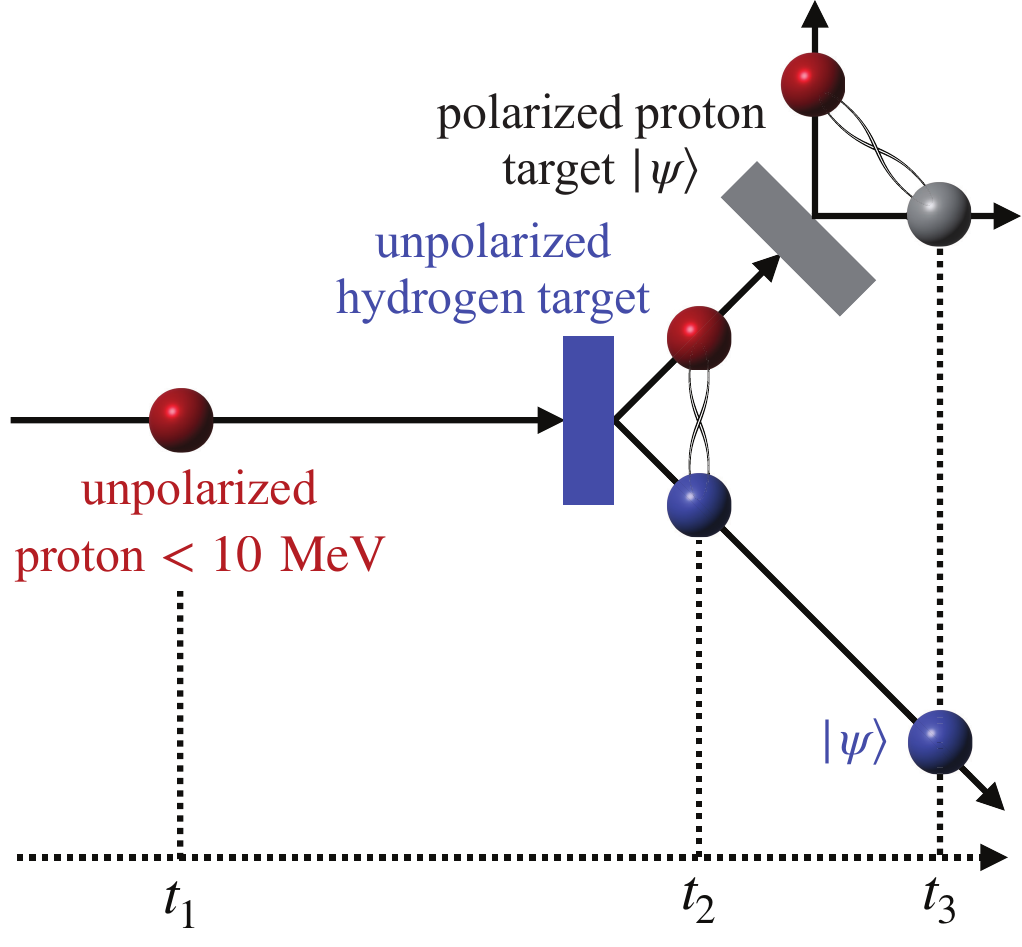}
	\caption{Schematic of the proton spin teleportation protocol at $E<10$~MeV. An unpolarized proton (red) with kinetic energy below 10~MeV scatters off an unpolarized liquid hydrogen target (blue rectangle) at time $t_1$, generating a spin-entangled proton pair via $S$-wave-dominated scattering. At time $t_2$, one of the scattered protons (red) interacts with a polarized proton target (gray) carrying the unknown spin state $\ket{\psi}$. The second low-energy scattering projects the red and gray protons onto the Bell-singlet state, completing the teleportation. At time $t_3$, the remote blue proton carries the teleported spin state $\ket{\psi}$.}
	\label{fig:qt_lowE}
\end{figure}

\textit{Setup}---The protocol involves three protons:
\begin{itemize}[leftmargin=*]
	\item Target proton (gray): The polarized target proton carrying the unknown spin state $\ket{\psi}_1 = \alpha\ket{0}_1 + \beta\ket{1}_1$ to be teleported.
	\item Incident proton (red): An unpolarized incident proton with kinetic energy below 10~MeV.
	\item Receiver proton (blue): An unpolarized proton in a liquid hydrogen target, which will receive the teleported state.
\end{itemize}
Unlike the 151~MeV protocol, this scheme does not require a pre-prepared entangled pair. Instead, the entanglement is generated \textit{in situ} through the first scattering event. A schematic illustration of the protocol is shown in Fig.~\ref{fig:qt_lowE}.

\textit{Spin state at $t_0$}---Initially, proton 1 carries the unknown spin state while protons 2 and 3 are completely unpolarized:
\begin{equation}
	\rho(t_0) = \ket{\psi}_1\bra{\psi}_1 \otimes \frac{I}{2} \otimes \frac{I}{2},
\end{equation}
where $I/2$ denotes the maximally mixed single-qubit state.

\textit{Spin state at $t_1$}---At time $t_1$, the unpolarized incident proton in mode 2 scatters with the unpolarized target proton in the liquid hydrogen target; we postselect events where the two outgoing protons populate two distinct arms, which we then label as modes 2 and 3. At energies below 10~MeV, $S$-wave scattering dominates, and the spin amplitude takes the form
\begin{equation}
	M_{23} \propto \ket{\Psi^-}_{23}\bra{\Psi^-}_{23}.
\end{equation}
Applying $(I_1 \otimes M_{23})$ to the initial state and normalizing, the post-scattering state becomes
\begin{equation}
	\ket{\Psi(t_1)} = \ket{\psi}_1 \otimes \ket{\Psi^-}_{23},
\end{equation}
where protons 2 and 3 are now maximally entangled in the Bell-singlet state. This entanglement generation through low-energy $pp$ scattering was first exploited in the nuclear Bell test by Lamehi-Rachti and Mittig in 1976.

Expanding this state in the Bell basis for qubits 1 and 2:
\begin{align}
	\ket{\Psi(t_1)} &= \frac{1}{2}\Big[\ket{\Phi^+}_{12} \otimes \left(\alpha\ket{1}_3 - \beta\ket{0}_3\right) + \ket{\Phi^-}_{12} \otimes \left(\alpha\ket{1}_3 + \beta\ket{0}_3\right) \nonumber\\
	&\quad + \ket{\Psi^+}_{12} \otimes \left(-\alpha\ket{0}_3 + \beta\ket{1}_3\right) + \ket{\Psi^-}_{12} \otimes \left(-\alpha\ket{0}_3 - \beta\ket{1}_3\right)\Big].
	\label{eq:lowE_t1}
\end{align}

\textit{Spin state at $t_2$}---At time $t_2$, the proton in the selected outgoing arm (mode 2) is directed to scatter with the polarized target proton in mode 1, again at low energy where $M_{12}\propto\ket{\Psi^-}_{12}\bra{\Psi^-}_{12}$. This operator projects onto the $\ket{\Psi^-}_{12}$ component of Eq.~(\ref{eq:lowE_t1}):
\begin{equation}
	(M_{12}\otimes \mathbf{1}_3)\ket{\Psi(t_1)} \propto \ket{\Psi^-}_{12} \otimes \left(-\alpha\ket{0}_3 - \beta\ket{1}_3\right).
\end{equation}
After normalization, the final state is
\begin{equation}
	\ket{\Psi(t_2)} = \ket{\Psi^-}_{12} \otimes \left(-\alpha\ket{0}_3 - \beta\ket{1}_3\right).
\end{equation}
The remote proton 3 now carries the spin state $-(\alpha\ket{0}_3 + \beta\ket{1}_3)$, which is identical to the original unknown state $\ket{\psi}_1$ up to an irrelevant global phase.

\textit{Comparison with the 151~MeV protocol}---The low-energy protocol differs from the 151~MeV protocol in several key aspects:
\begin{enumerate}[leftmargin=*]
	\item \textit{Entanglement source}: The low-energy protocol generates entanglement on-the-fly through the first $pp$ scattering, whereas the 151~MeV protocol requires a pre-prepared entangled pair from, e.g., the $p(d,{}^{2}\mathrm{He})n$ reaction.
	\item \textit{Scattering operator}: At $E<10$~MeV, the amplitude $M\propto\ket{\Psi^-}\bra{\Psi^-}$ acts as a projection operator onto the Bell-singlet state. At 151~MeV and $90^\circ$, the amplitude $M\propto\ket{\Psi^+}\bra{\Phi^-}$ acts as a Bell transition operator connecting different Bell states.
	\item \textit{Final state transformation}: In the low-energy case, the teleported state matches the original state exactly (up to a global phase). In the 151~MeV case, the teleported state differs by an interchange $\ket{0}\leftrightarrow\ket{1}$, corresponding to an inversion of the quantization axis.
\end{enumerate}
Both protocols implement probabilistic teleportation: successful events are heralded by the detection of the scattered proton pair at the appropriate kinematics.

\newpage

\bibliography{references}